# Ultralow-Power Single-Sensor-Based E-Nose System Powered by Duty Cycling and Deep Learning for Real-Time Gas Identification

*Taejung Kim[‡, 1], Yonggi Kim[‡, 2], Wootaek Cho[1], Jong-Hyun Kwak[1], Jeonghoon Cho[2], Youjang Pyeon[2], Jae Joon Kim[\*, 2], Heungjoo Shin[\*, 1].*

[1]Department of Mechanical Engineering, Ulsan National Institute of Science and Technology (UNIST), Ulsan, 44919, Republic of Korea

[2]Department of Electrical Engineering, Ulsan National Institute of Science and Technology (UNIST), Ulsan, 44919, Republic of Korea

[‡] These authors contributed equally

*Corresponding Author Email: J. J. Kim (jaejoon@unist.ac.kr), H. Shin (hjshin@unist.ac.kr)

ABSTRACT

This study presents a novel, ultralow-power single-sensor-based electronic nose (e-nose) system for real-time gas identification, distinguishing itself from conventional sensor-array-based e-nose systems whose power consumption and cost increase with the number of sensors. Our system employs a single metal oxide semiconductor (MOS) sensor built on a suspended 1D nanoheater, driven by duty cycling—characterized by repeated pulsed power inputs. The sensor's ultrafast thermal response, enabled by its small size, effectively decouples the effects of temperature and surface charge exchange on the MOS nanomaterial's conductivity. This provides distinct sensing signals that alternate between responses coupled with and decoupled from the thermally enhanced conductivity, all within a single time domain during duty cycling. The magnitude and ratio of these dual responses vary depending on the gas type and concentration, facilitating the early-stage gas identification of five gas types within 30 s via a convolutional neural network (classification accuracy = 93.9%, concentration regression error = 19.8%). Additionally, the duty-cycling mode significantly reduces power consumption by up to 90%, lowering it to 160 μW to heat the sensor to 250 °C. Manufactured using only wafer-level batch microfabrication processes, this innovative e-nose system promises the facile implementation of battery-driven, long-term, and cost-effective IoT monitoring systems.

INTRODUCTION

The growing demand for cost-effective and low-power gas sensors is fueled by the rapid expansion of wireless applications in various monitoring platforms, such as stationary, drone-based, and wearable gas monitoring systems. More recently, their applications have expanded to include areas such as home indoor air quality monitoring, early disease diagnosis through exhalation analysis, and food spoilage monitoring.[1,2] Among various sensing materials, metal-oxide semiconductors (MOS) have been chosen as the most prominent choice for gas sensors in wireless sensor networks owing to their cost-effective synthesis, material design compliance, high sensitivity, and suitability for sensor miniaturization.[3,4] MOS gas sensors detect the target gas simply by measuring the change in electrical resistance upon exposure to the target gas. The major challenges encountered by MOS-based gas sensors, such as their inherent high power consumption and poor selectivity, have restricted their real-time wireless applications.[5,6] The primary contributor to high power consumption in MOS-based gas sensors is the need for activation energy, which is required for promoting surface molecular adsorption and desorption of the target gases on the metal oxide surface. Various techniques have been developed to minimize power consumption in these sensors, including sensor operation with a miniaturized joule heater, using room-temperature sensing materials, or through photoactivation.[7–10] Despite significant power reduction, room temperature sensing suffers from poor recovery rates.[8] This issue arises because its thermal energy is insufficient to effectively desorb the target gases from the sensing sites. Meanwhile, Joule-heating methods promote gas desorption by heating the sensing material to temperatures between 100 and 400 °C.[11] Furthermore, the MOS sensor output is influenced by two main factors: the change in the MOS conductivity due to temperature variation and the surface charge exchange caused by the target gas; the latter is known as chemisorption. Consequently, these two factors result in a coupled sensor output signal. Therefore, to ensure stable detection through uniform heating of MOS materials at low power consumption, miniaturized heater structures are designed as long aspect-ratio suspended bridges, providing effective thermal insulation from the substrate.[12–15]

In addition, from the viewpoint of sensor operation, duty cycling has been rigorously explored as an alternative power-saving strategy for heater-driven MOS gas sensors.[16–19] As indicated by the yellow curve illustrated in **Figure 1**a, the heater is periodically toggled between power-on and low-power (or power-off) states in the duty-cycling mode, minimizing the active operation time of the heater. Therefore, the average power consumption is effectively reduced in proportion to the power-off time ($\tau_s$, or sleep time) compared to that in the constant power supply mode. Hence, reducing the duty cycle—defined as the ratio of the wake-up period ($\tau_w$) to the total cycle period ($\tau$)—is the key to power saving in duty-cycling mode. However, the switching kinetics of the sensor's heater are typically not immediate, as indicated by the black dotted curve shown in **Figure 1**a. This means that sensors require a certain duration for thermal stabilization (or thermal time constant) in both the heating and cooling periods. Thus, the actual sensor operation time is reduced compared to heater-on time, leading to wasted energy. Therefore, rapidly responsive heaters are required to minimize the duty cycle. The thermal time constant ($\tau_T$), defined as the time required for the change in the heater temperature to reach 63.2% of the difference between the initial and final stabilized temperatures during a sudden power input, is given by

$$\tau_T = \frac{\rho V C}{h A_s} \tag{1}$$

where $\rho$ is the density, $V$ is the volume, $C$ is the specific heat, $h$ is the heat transfer coefficient, and $A_s$ is the surface area of the body. In this regard, the thermal time constant decreases as the body size decreases, accelerating the development of MOS gas sensors built on miniaturized heaters.[20,21] However, the reduction to micrometer-sized heaters is not sufficient for achieving duty cycles below 20%.[16–21] Their thermal time constants are tens of milliseconds range in most cases.

Another major challenge faced by MOS-based gas sensors is their gas cross-sensitivity or poor selectivity. For instance, MOS sensors typically respond to multiple gas types, hindering accurate gas identification. Various approaches have been employed to overcome the poor selectivity of MOS-based sensors. These include combining sensors into arrays,[22–24] improving sensor response through temperature and power modulation,[25–28] extracting and analyzing complex features from sensor data,[29–31] and employing broad-range impedance spectroscopy.[32] These approaches expand the dimensionality of sensor response, thereby enhancing the selectivity of sensors. For example, an array of S different sensors operating in M different selection modes yields P = MS parameters or data channels.[33] Nonetheless, they often require complex data acquisition processes and multiple sensors for accurate prediction, limiting real-time recognition. For further enhancing selectivity, electronic nose (e-nose) systems have emerged as a promising solution. These systems identify gases by analyzing characteristic gas responses from multiple sensors through advanced machine learning (ML) algorithms.[23,31] Electronic nose (e-nose) systems, employing a gas sensor array, to enhance selectivity for accurate gas identification, paradoxically lead to increased power consumption and costs proportionate to the number of sensors used. Thus, researchers grapple with a persistent trade-off: enhancing prediction accuracy while managing practical limitations such as energy efficiency and system scalability.

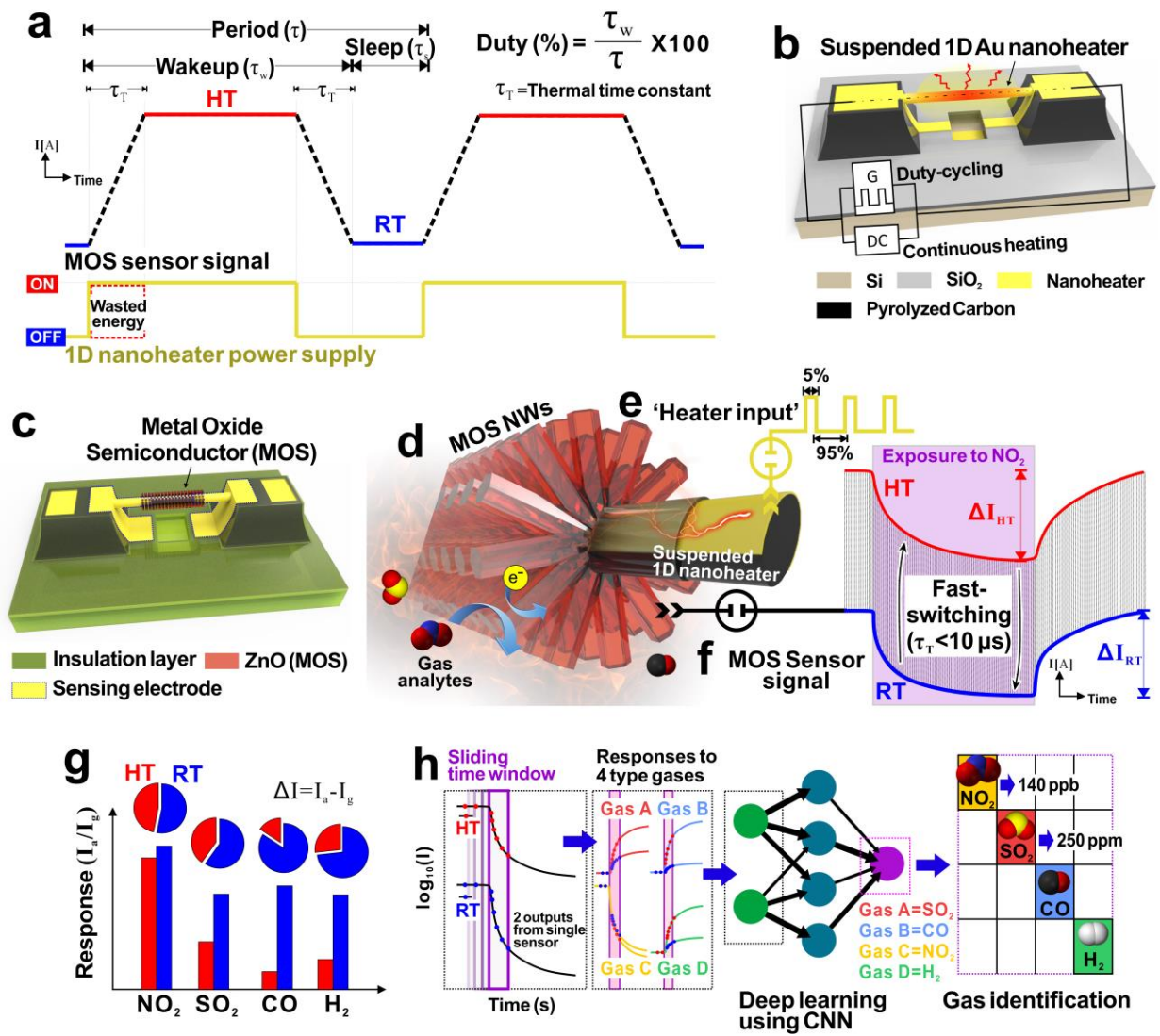

**Figure 1**. (**a**) Schematic of sensor temperature change under the duty-cycling mode [yellow curve: power input; black dotted curve: current signal during the temperature transition; red curve: current signal at high temperature (HT); blue curve: current signal at room temperature (RT)]. (**b–d**) Schematic of a MOS gas sensor based on a suspended 1D nanoheater: (**b**) suspended 1D nanoheater; (**c**) gas sensor constructed by regioselectively growing a MOS nanowire network in the central region of the 1D nanoheater; (**d**) cross-sectional view of the MOS NW network grown circumferentially around the 1D nanoheater; schematic representation of (**e**) the repeated pulsed power and (**f**) the corresponding transient current change of the MOS sensing site under duty

cycling. (**g**) Schematic chart of dual gas responses at HT and RT under duty-cycling mode. (**h**) Schematic of machine-learning-based gas identification using dual-response signals from a single sensor.

In this study, we demonstrate a significant breakthrough in the field of MOS-based gas sensors by addressing both their high power consumption and limited selectivity, which is achieved through the operation of a single nanoheater-based MOS gas sensor in duty-cycling mode. Our sensors, featuring a MOS nanowire (NW) network radially grown on a suspended 1D nanoheater (**Figures 1**b–d), were manufactured through wafer-level microfabrication processes, as reported in our previous studies.[34,35] The heater's nano-sized body and high aspect ratio (width ~300 nm, thickness ~500 nm, length = 130 μm), regioselectively grown MOS NWs in the central region of the heater, and robust thermal insulation (due to the suspended high-aspect-ratio architecture) enabled instantaneous heating of the MOS sensing site at less than 2 mW. The remarkably small thermal time constant of the nanoheater, clocking less than 10 μs, dramatically accelerates the wake-up of the sensor in a power duty. This enabled a significant reduction in the duty cycle, diminishing power consumption to less than 160 μW, which translates to 90% energy savings compared to consumption under the constant power mode. Furthermore, we introduce the gas identification capability implemented by a 'single' sensor, owing to its unique attribute of 'dual gas responses' induced by duty cycling. In addition to gas response at high temperature (HT) under the heater power-on state, distinct and transient gas sensing behavior, decoupled from the thermally enhanced electrical conductivity of the MOS, was also observed even in the room temperature (RT) mode (power-off mode) in a cycled power input (**Figures 1**e and f). This is due to a significantly faster thermal response compared to gas adsorption/desorption at the MOS surface. Therefore, a single sensor can generate dual responses without requiring complex multi-sensor setups and the corresponding increased power demands (**Figure 1**g). These responses, both coupled with and decoupled from the thermally enhanced conductivity, were measured along a single time domain and analyzed by a convolutional neural network (CNN), leading to real-time gas identification (**Figure 1**h). After rigorous training and validation, the ML model demonstrated high accuracy in predicting gas types and concentrations. With an accuracy rate surpassing 90%, the CNN model effectively distinguished between five types of gases (air, $NO_2$, $SO_2$, CO, and $H_2$) based on their distinct responses at HT and RT. In addition, the mean absolute percentage error (MAPE) of the concentration predicted through the regression model was calculated as only approximately 20%. This single-sensor-based gas identification method will contribute to advancing energy-efficient and cost-effective e-nose systems with excellent discriminatory capabilities.

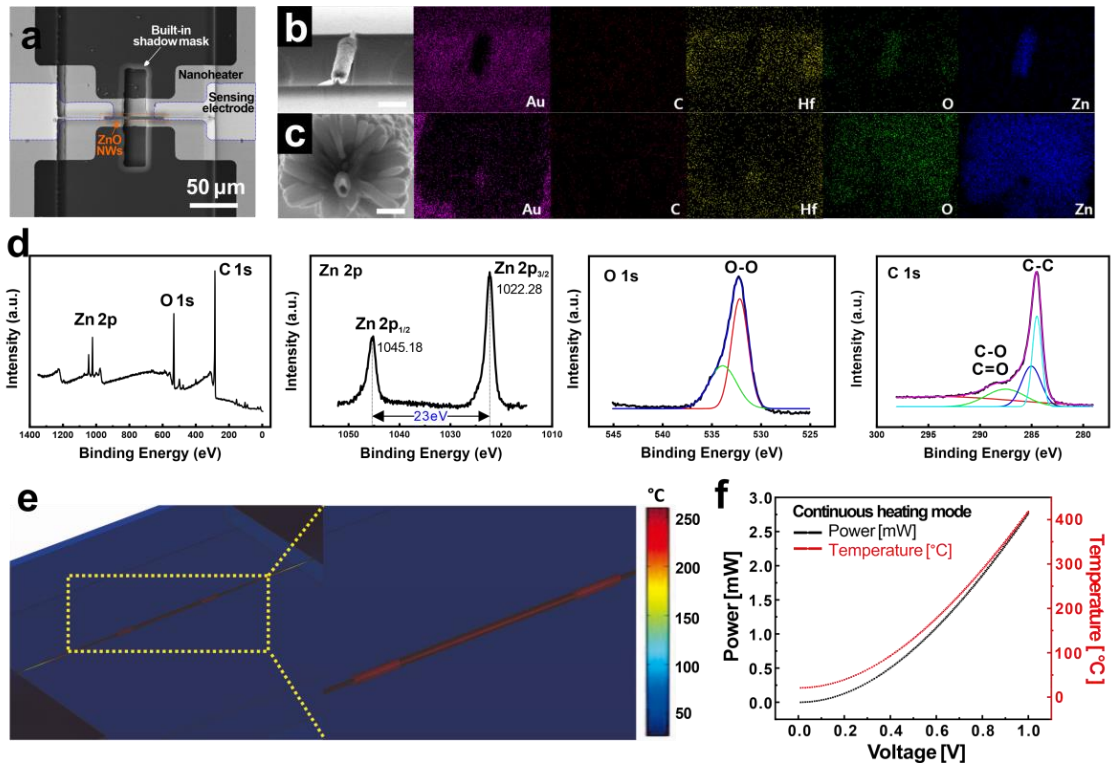

**Figure 2**. Morphological, microstructural, and thermal characteristics of a MOS gas sensor based on a suspended 1D nanoheater: (**a**) Top view (SEM image) of a ZnO NW network grown in the central region of the 1D nanoheater; (**b, c**) bird's eye view (SEM images) of the ZnO NW network circumferentially grown on the nanoheater (inset scale bars: (**b**) 3 μm, (**c**) 300 nm), along with corresponding EDS mapping images; (**d**) XPS spectra of a ZnO NW network for an overall composition, Zn 2p region, O 1s region, and the C 1s region, listed from the left to right; (**e**) simulated temperature profile of the suspended 1D nanoheater coated with a MOS NW network (heater power = 2 mW); (**f**) Experimentally measured temperature and power consumption of the suspended 1D nanoheater-based MOS gas sensor according to the applied voltage to the nanoheater in continuous heating mode.

RESULTS AND DISCUSSION

**Fabrication of a Gas Sensor Platform Based on a MOS NW Network Built on a Suspended 1D Nanoheater.** The presented sensor comprises a MOS NW network in the central region of a suspended 1D nanoheater, as illustrated in **Figure 2**a. In this study, the suspended 1D nanoheater-based gas sensors were fabricated following the previously reported method.[34] As shown by the schematic fabrication steps in **Figure S1**, the 1D nanoheater in a suspended architecture was fabricated *via* selective metal coating on a suspended carbon NW backbone. Subsequently, a MOS NW network was grown radially along the nanoheater's central region (**Figure S2**), where the heater temperature rises uniformly. Despite the complex 3D mixed-scale architecture of the sensor platforms, all the fabrication steps were based on wafer-level batch microfabrication processes, ensuring cost-effective manufacturing. A more detailed description of the fabrication steps is presented in the Methods section. As the nanoheater backbone, a suspended carbon nanowire was fabricated using the carbon-microelectromechanical system (C-MEMS) technology, which enables the facile fabrication of micro/nano carbon structures by pyrolyzing pre-patterned polymer structures. In this study, a suspended polymer micro-sized wire (width ~1.3 μm, thickness ~2.5 μm, length ~130 μm, wire-to-substrate separation ~20 μm) was fabricated *via* two successive photolithography steps (first, UV exposure at high energy for the polymer post structures; second, UV exposure at low energy for the suspended polymer wire). Subsequently, the polymer wire was carbonized through a pyrolysis process in a vacuum. Most elements, except carbon, escape from the polymer during pyrolysis, resulting in a volume reduction of up to 90%. This volume reduction is facilitated by smaller structure sizes and more efficient mass transfer, leading to a greater reduction in volume for smaller suspended structures.[36] Consequently, the suspended micrometer-sized polymer wire was converted into a nanoscale carbon wire (width ~250 nm, thickness ~300 nm). The resultant pyrolyzed carbon shows superior mechanical properties, sufficient to withstand wet microfabrication processes.[37] In this study, to make the suspended carbon nanowire work as the backbone of the nanoheater, its conductivity was tuned sufficiently low so that most of the current flowed through the metal heater layer coated on the carbon wire. This conductivity control was facilitated by adjusting the pyrolysis temperature.[38] After forming the carbon backbone, a 50-nm-thick Au layer, a heater electrode material, was deposited selectively at the 1D nanoscale backbone using microscale patterning processes. This was enabled by a unique built-in shadow mask structure consisting of $SiO_2$ eaves protruding from the top edges of an isotropically etched Si trench, as shown in the inset of **Figure S2**f. When a metal layer is deposited on top of the built-in shadow mask using highly directional deposition methods such as evaporation, the metal layer is disconnected at the undercut below the eaves. Consequently, a continuous metal coating between two carbon posts is made solely along the suspended carbon backbone as long as the photoresist mask width is smaller than the built-in shadow mask, allowing nanopatterning using conventional photolithography, as illustrated in **Figure S3**. The built-in shadow mask was formed prior to the C-MEMS process. The detailed working principle of the built-in shadow mask was reported elsewhere.[34,35] Subsequently, an insulation layer was deposited on the entire substrate and then etched for the electrical connection to the nanoheater. The MOS NWs (ZnO NWs in this study) were grown using the hydrothermal method, which grows nanostructures from a seed layer in a heated growth bath. In this study, the MOS NWs were locally grown in the heater's central region by heating the entire substrate. This was enabled by patterning a seed layer in the heater's central region *via* conventional photolithography and coating processes. The suspended 1D nanoheater maintained physical integrity during the NW patterning processes due to the excellent mechanical robustness of the carbon backbone, as mentioned earlier.

**Characterization of the Gas Sensing Material. Figures S2**b and c show scanning electron microscopy (SEM) and the corresponding energy dispersive spectroscopy (EDS) mapping images of ZnO NWs (diameter: ~100 nm, length: 0.5–1 μm) grown in the central region of the 1D nanoheater. The nanoheater was fractured at its center, revealing the prominent morphology of the NW network distributed circumferentially around the nanoheater. The EDS mapping results help clearly identify the mass distributions of the sensor and heater materials, such as pyrolyzed carbon, Au, $HfO_2$, and ZnO. In the magnified views (**Figure 2**c), Au and Hf elements are mainly observed on the carbon nanowire surface, whereas Zn and O are predominantly distributed at the MOS NW network. The composition and chemical states of ZnO NWs were analyzed using X-ray photoelectron spectroscopy (XPS) (**Figure 2**d). The ZnO NW sample for the XPS test was prepared on top of a pyrolyzed carbon film through the same process used for the gas sensor. The high-resolution Zn 2p spectrum shows two peaks at 1022.28 and 1045.18 eV, corresponding to Zn 2p$_{3/2}$ and Zn 2p$_{1/2}$, respectively. In addition, the energy difference of approximately 23 eV between these two peaks confirms that Zn exists primarily in the Zn$^{2+}$ chemical state. The C1s spectrum comprises three peaks at 284.5, 285.1, and 287.5 eV. The peak at 284.5 eV corresponds to aromatic and aliphatic carbons (C–C), and the peaks at 285.1 and 287.5 eV represent carbon bound to oxygen (C–O and C=O, respectively).

**Evaluation of 1D Nanoheater Operation.** Suspended 1D nanostructures can be heated using ultralow power because of their small size and excellent thermal insulation. However, a 1D nanoheater is mostly heated in the central region because of its high aspect ratio, resulting in a parabolic temperature distribution along its longitudinal direction. This uneven heating hinders the gas detection capabilities of the MOS-based gas sensor because its gas sensitivity varies significantly depending on the working temperature. To address this issue, we selectively grew a MOS NW network in the central region of a suspended 1D nanoheater, where the temperature rises the most. This regioselective localization ensured uniform heating of the sensing site, with a temperature variance of less than 2% compared to the average temperature, as indicated by the simulated temperature profile shown in **Figure 2**e. This uniform heating capability was experimentally confirmed by comparing the changes in resistance of the ZnO NW network upon exposure to a gas analyte when heating the entire sensor chip using a hotplate and when locally heating the ZnO NWs using the nanoheater, as described in our previous study.[34] Owing to the small size and suspended architecture, this sensor platform featured superior power efficiency (138.8 K mW$^{-1}$), expressed by temperature rise according to the heater power (*i.e.,* a 225 °C rise by 1.62 mW, as shown in **Figure 2**f). This experimental result was in good agreement with the simulation result. The power consumption in the continuous power mode of this nanoheater-based sensor platform is just ~4% of the total power required for Bluetooth communication typically used in wireless monitoring systems. Moreover, further power reduction can be achieved by implementing a duty-cycling mode enabled by the fast thermal response of the proposed sensor. Experimental results on the effect of the duty-cycling mode on sensor performance are discussed in the subsequent section.

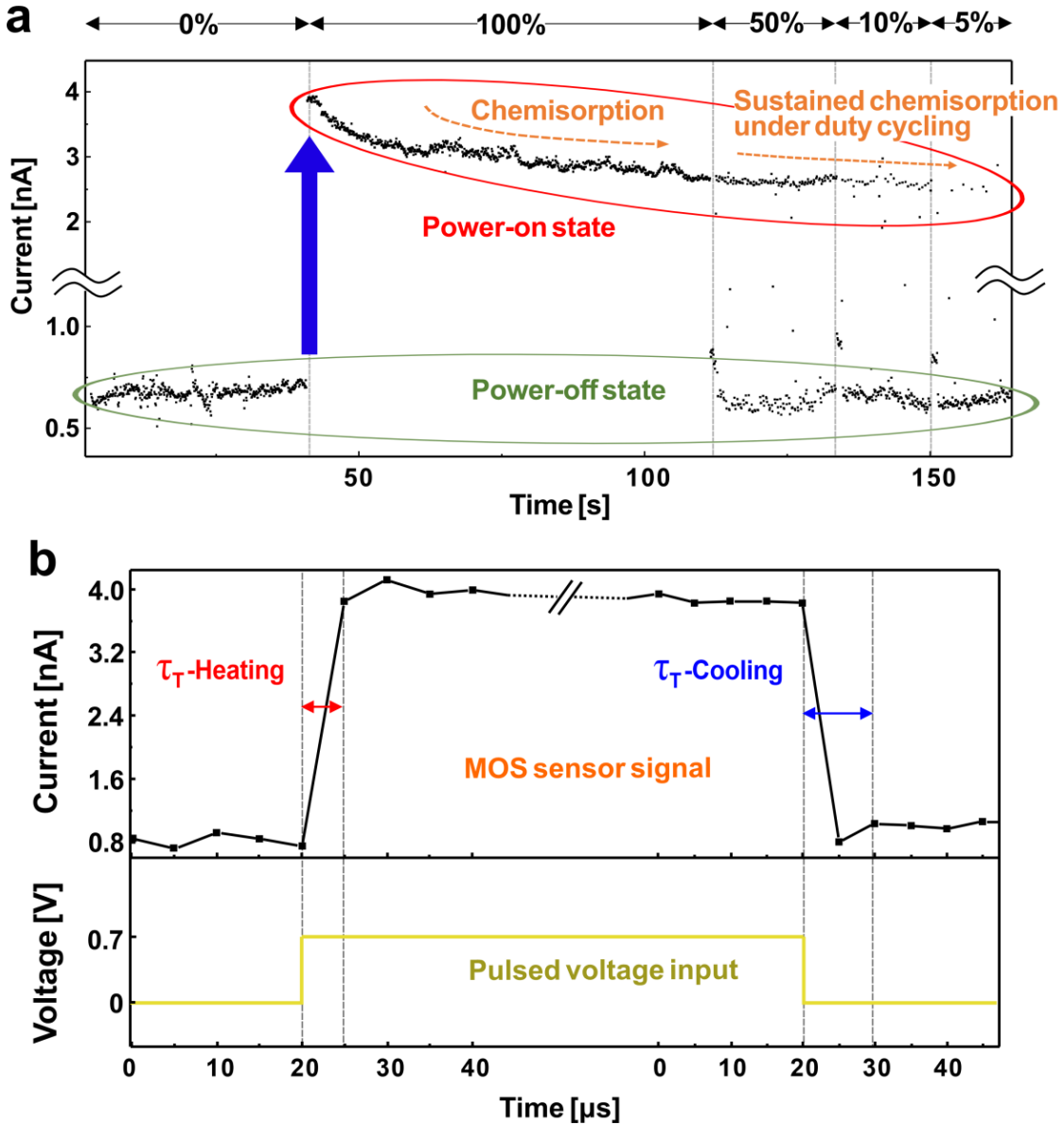

**Figure 3**. Switching kinetics of the 1D nanoheater-embedded MOS gas sensor in duty-cycling mode: (**a**) Real-time current signal in dry air from the MOS NW network under varying duty cycles

(100%, 50%, 10%, and 5% at 1 Hz) (dots inside the red oval: data points in the power-on state; dots inside the green oval: data points in the power-off state; blue arrow: initiation of heater operation). (**b**) Measurement of the MOS NW's thermal time constants during heating and cooling of the duty-cycled 1D nanoheater (top diagram: current signal through the ZnO NW network, bottom diagram: voltage input to the nanoheater).

**Characterization of the Duty-Cycling Effect on Sensor Signals.** In duty-cycling mode, power consumption is determined either by the increase in heater temperature or by the duration of the power-on state (or duty). Therefore, power saving can be achieved by reducing the duty at a fixed heater temperature. However, there are limits to reducing the duty cycle because the amount of data that can be obtained in a short period of time is limited by the measurement equipment. As the duty cycle is reduced, the number of readout data points in the power-on state decreases, as demonstrated in **Figure S4**. When the duty cycle is minimized to its limit, despite actual heating occurring, the data acquisition speed cannot keep pace with the duty-cycling speed, rendering it unmeasurable. In this study, the heating frequency was adjusted sufficiently large (1 Hz) to ensure a data acquisition time of 60 ms, allowing the equipment to obtain at least one data point in the heated state, even at the smallest duty cycle (0.06%). **Figure 3** shows the current signal from a ZnO NW network under nanoheater operation in duty-cycling mode in dry air. Effective duty cycling-based sensor operation requires rapid response and long-term durability of the heater because the sensor's response depends on the nanoheater temperature. Immediately after the onset of heating, a fivefold increase in the conductivity of the metal oxide was observed, as indicated by the blue arrow in **Figure 3**a. This is attributed to the negative temperature coefficient of resistance of the semiconductor material. **Figure 3**b shows an enlarged view of the rapid current-signal changes during heating and cooling, indicating thermal time constants of approximately 5 and 10 µs for heating ($\tau_{\text{T-Heating}}$) and cooling ($\tau_{\text{T-Cooling}}$), respectively. These rapid thermal responses ensure stable sensor operation in duty-cycling mode at small duties and high frequencies, enabling real-time monitoring while simultaneously saving power. After the current surge, the current signal gradually stabilized in constant power mode (duty cycle = 100%), as shown in **Figure 3**a. This stabilization is attributed to the relatively slow chemisorption of $O_2$—the phenomenon where oxygen ions adhere to the MOS surface through the reduction of $O_{2(gas)}$ ($O_{2(gas)} + e^- \rightarrow O_2^-{}_{(ads)}$), increasing the resistance of *n*-type MOS. This chemisorption was sustained even under duty-cycled power modes (duty cycle = 50%, 10%, and 5%), as indicated by the data dots in the heating state (or power-on state) in **Figure 3**a. Furthermore, when the heater power was switched to a power-off state in duty-cycling mode, the sensor's resistance was slightly higher than that measured before the heater operation. This indicates the persistence of the chemisorbed oxygen ions even in the power-off state. In addition, the recurring stabilized resistance signals at both the power-on and power-off states suggest that the temperature changes in duty-cycling mode are sufficiently rapid to prevent the desorption of oxygen ions from the MOS surface. These persistent reactive oxygen species on the MOS surface facilitate the measurement of changes in resistance, which are decoupled from the thermally-coupled changes in electrical conductivity of the semiconducting material. The related gas-sensing results will be discussed in the next section.

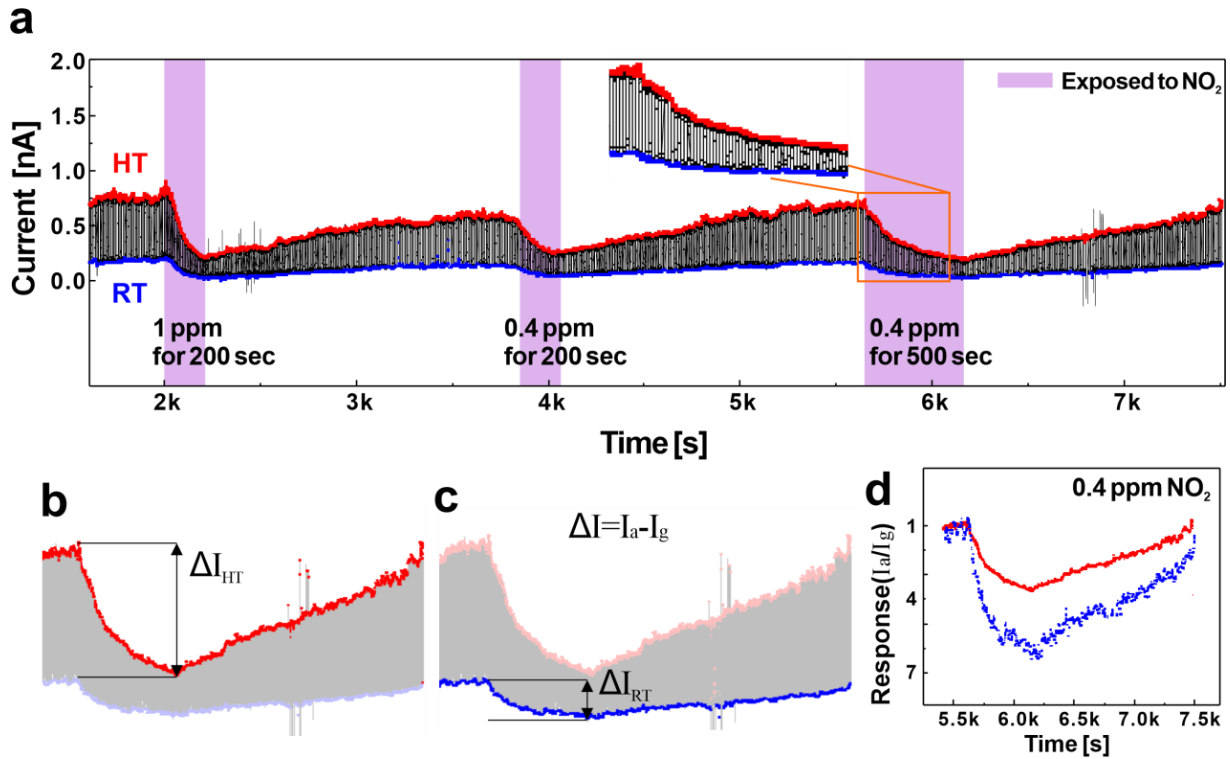

**Figure 4**. Transient current signal upon recurrent exposure to $NO_2$ (0.4–1 ppm) in duty-cycling mode (duty cycle = 60%, frequency = 2 Hz, heater power = 840 μW): (**a**) All data points (inset: magnified data set); (**b, c**) current data at (**b**) HT (red dots) and (**c**) RT (blue dots); (**d**) comparative graph of the sensor's responses at HT and RT. The red and blue dots represent data at the 90% and 10% percentiles in a single duty cycle, respectively. The purple-shaded regions in the graph in Figure (**a**) represent the gas injection periods.

**Evaluation of Effect of Duty Cycling on Gas Response.** The chemisorbed oxygen ions ($O_2^-$ $_{(ads)}$) bound to the MOS surface react with the oxidizing and reducing gas analytes, leading to changes in the electrical resistance of the sensing material. To investigate the sensor's responsiveness and conductive behavior for different gas environments in duty-cycling mode, a comprehensive analysis was conducted. First, the gas response to $NO_2$ of the proposed 1D nanoheater-embedded gas sensor was tested in duty-cycling mode, as shown in **Figure 4**. Upon exposure of the *n*-type metal oxide surface (*e.g.,* ZnO) to an oxidizing gas such as $NO_2$, electrons at the MOS are extracted through gas chemisorption, causing a decrease in the conductivity of the MOS. In this event, $NO_2$ mainly contributes to the increase in MOS resistance owing to its higher electron affinity compared to that of $O_2$. The gas response for oxidizing gases ($NO_2$) is defined as $I_a/I_g$ (where $I_a$ and $I_g$ represent the sensor resistance in air and upon exposure to oxidizing gas, respectively). Conversely, the response pattern is reversed when the *n*-type semiconductor-based sensor interacts with reducing gases such as CO, $SO_2$, and $H_2$. Therefore, in this case, the gas response is defined as $I_g/I_a$. In chemisorption, the target analytes require sufficient energy to activate the process, which is typically provided by heating the MOS materials. Consequently, the change in MOS conductivity upon exposure to gas analytes is coupled with the thermally enhanced conductivity. However, in duty-cycling mode, these chemisorption-based gas responses are observed not only in the power-on state but also in the power-off state, as shown in **Figure 4**a.

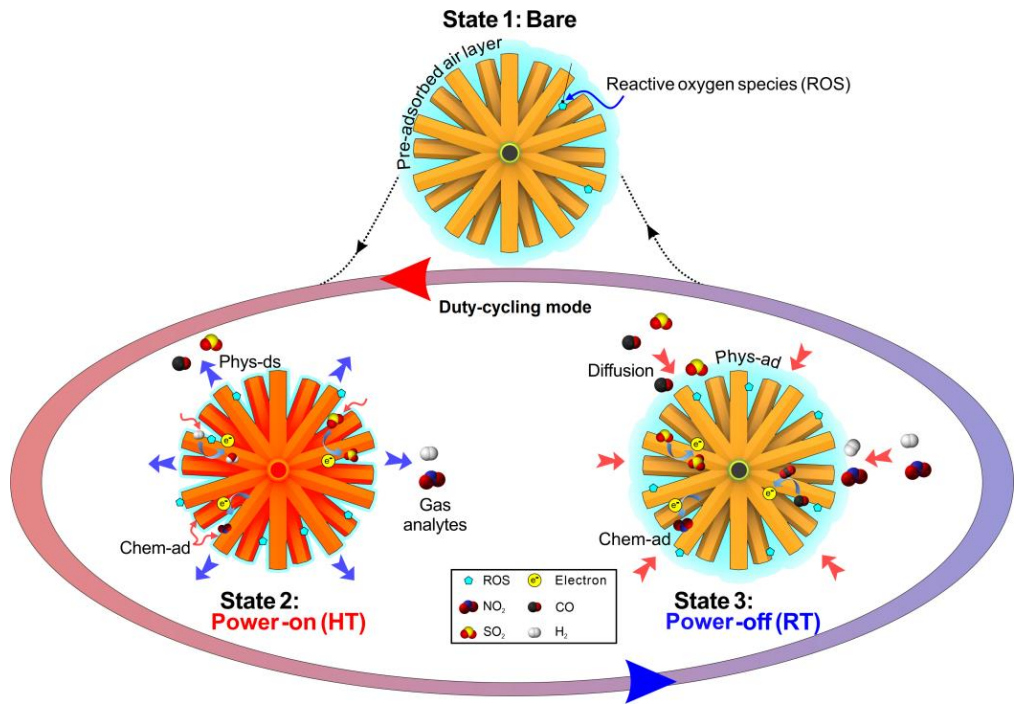

**Figure 5**. Schematic of the mechanism for physisorption/chemisorption of gas analytes onto circumferentially grown MOS NWs in duty-cycling mode. ('ad': adsorption, 'ds': desorption, 'chem-ad': chemisorbed species, 'phys-ad': physisorbed species)

To gain insight into the gas response in the power-off state in duty-cycling mode, it is crucial to consider the impacts of duty-cycle-induced temperature modulation on the adsorption, redox, chemisorption, and desorption of gas analytes. In this study, ZnO NWs grown circumferentially on the 1D nanoheater render the MOS surfaces more accessible to gas molecules, and the suspended architecture facilitates efficient gas transfer from the bulk. As shown in the diagram in **Figure 5**, the target gas molecules diffuse and condense on the MOS surface during the power-off state (RT: States 1 and 3) and react with the adsorbed oxygen molecules during the power-on state (HT: State 2).[17] In detail, before initiating duty cycling (State 1), the MOS surfaces are predominantly covered by a thick pre-adsorbed air layer (consisting of a thin chemisorbed layer and a thick physisorbed layer).[39] This layer hampers the direct interaction between the analyte molecules and the MOS surface, leading to negligible electrical responses.[17,] When the MOS is heated *via* duty-cycling (State 2), this process facilitates the chemisorption of the analyte and oxygen by providing sufficient activation energy.[17,19,40] During the chemical adsorption of the gas analytes, electrons are supplied to oxidizing analytes or withdrawn from reducing species, resulting in changes in the MOS conductivity. Meanwhile, this process is accompanied by the partial desorption of the analytes. When the temperature of the MOS is modulated down intermittently *via* duty cycling (State 3), most of the pre-chemisorbed molecules are trapped within a potential well.[40] This entrapment occurs because their desorption requires a relatively high activation energy.[41] Therefore, this intermittent cooling fosters the accumulation of chemisorption.

Thus, the duty-cycling-driven gas sensor enables gas detection in the RT state by measuring the changes in resistance that are primarily due to chemisorption rather than the thermally induced changes in conductivity. Additionally, while both reactions at HT and RT were recorded through a single time domain, they exhibited different magnitudes of change, as shown in **Figures 4**b–d. In the RT state, both the current signal and the signal-to-noise ratio were smaller than those in the HT state, yet the sensor response was higher. This result can be attributed to (1) the entrapment of pre-chemisorbed gas analytes and reactive oxygen species in the potential well, and (2) the influence of temperature on both surface charge exchange and the electrical properties of the sensing channel. However, quantitatively describing temperature-dependent resistance is difficult due to diverse factors such as surface defect states, the nature and concentration range of the measured gas, sensing material characteristics (grain size, surface-to-volume ratio), dopants, interconnects between grains, and the geometry and material of sensing element electrodes.[32,42–44] Consequently, distinctive temperature-dependent dual sensing responses are exhibited even within a single gas detection, as indicated by the red (HT data or coupled data) and blue (RT data or decoupled data) dots in **Figure 4**. In this study, the HT and RT data sets were collected by filtering the top 90% and bottom 10% of all data in a single thermal cycle, respectively. These dual data patterns were consistently maintained during the duty-cycling process, resulting in a unique sensing response.

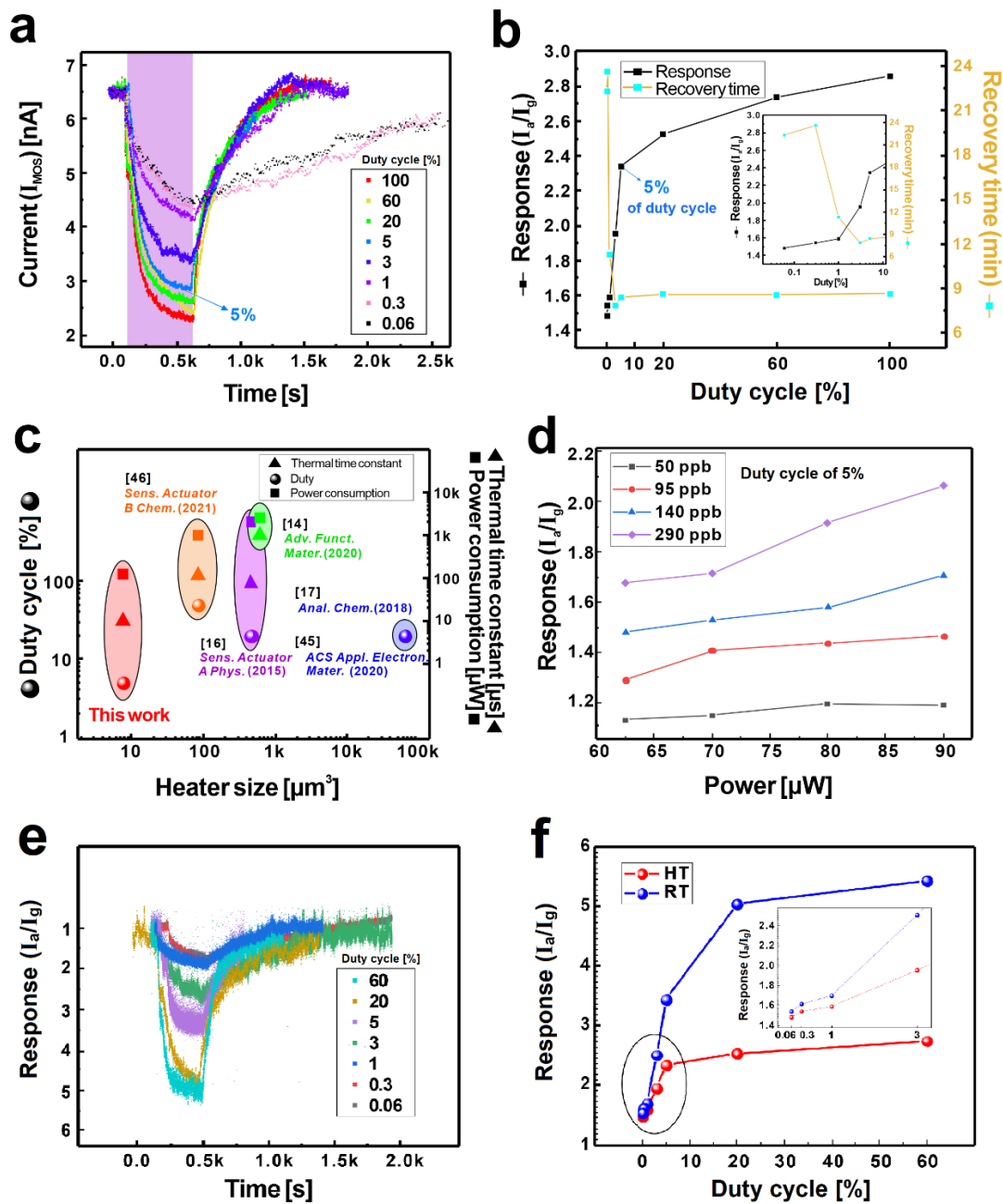

**Figure 6.** Gas responses of the 1D nanoheater-embedded sensor to 200 ppb $NO_2$ under various duty-cycling conditions: (**a**) Transient HT current signals at various duty cycles (0.06%–100%) (inset: magnified view); (**b**) gas responses and recovery times in the HT state corresponding to duty cycles (0.06%–100%); (**c**) comparison of the heater size, thermal time constant, duty cycle, and power consumption of the presented sensor in the HT state with other heater-integrated

sensors;[14,16,17,45,46] (**d**) gas responses at a 5% duty cycle under various $NO_2$ concentrations and power conditions in the HT state; (**e**) transient RT gas responses obtained under various duty-cycle conditions (60%, 20%, 5%, 3%, 1%, 0.3%, and 0.06%); (**f**) comparison between the RT (blue) and HT (red) responses under various duty cycles (inset: enlarged view of data points inside the black oval).

Based on the aforementioned feasibility test for the duty-cycling-based sensor operation, the effect of the duty cycle on the gas sensing performance was investigated. The gas responses at HT decreased with a reduction in the duty cycle despite the power savings achieved in duty-cycling mode, as shown in **Figures 6**a and b. Concurrently, the recovery time was extended with the reduction in the duty cycle. These adverse effects are attributed to the significantly slower kinetics of gas chemisorption/desorption on the MOS surfaces compared to the nanoheater's thermal response, as demonstrated by the gas response at RT state in duty-cycling mode in **Figure 4**. Thus, with a reduction in the duty cycle, the net time for the chemical reaction diminishes while desorption time increases, leading to a decrease in the gas response. This also applies to the desorption of the gas analyte from the MOS surfaces during the recovery period, where activation energy for desorption is required. These findings emphasize the importance of optimizing the duty cycle for the sensor's efficiency and responsiveness. As shown in **Figure 6**b, no substantial increase in the recovery time was found at a duty cycle of 5% compared to the greater duty cycles, whereas its response reached approximately 82.4% of the response obtained with a constant power mode (duty cycle = 100%). As the duty cycle was reduced to below 1%, the system encountered a stark 50% decrease in response, paired with a doubling of the recovery time. Despite this deterioration of the sensor capabilities, significant energy savings of up to 99% were achieved.

This suggests that the proposed sensor platform can continue monitoring by adjusting its duty cycle in power-restricted situations without substantially compromising detection performance.

The power-saving excellence of the proposed sensor was evaluated by comparing its duty, thermal time constant, and power consumption with those of other heater-integrated gas sensors. The presented sensor demonstrated superior heating efficiency, requiring 90 µW to heat the sensor to 250 °C at a duty cycle of 5%. This efficiency is the highest among recently published works, as presented in **Figure 6**c and **Table S1**.[14,16,17,45,46] This superior performance is attributed to the suspended 1D nano-sized heater and regioselective MOS NW integration, which enable ultralow power operation and ultra-fast power-switching capability (switching time of only 10 µs).

The sensitivity of the MOS-based sensors is closely correlated with their operating temperature. We assessed the temperature dependency of the proposed sensor's response in duty-cycling mode. The temperature of the sensing material was controlled by adjusting the amplitude of the current supplied to the heater at a duty cycle of 5%, a duty cycle that demonstrated a response comparable to that of continuous heating. The temperature of the MOS NW network was evaluated by comparing the changes in resistance observed when the entire sensor chip was heated using a hot plate with those observed when the 1D nanoheater was operated in constant power mode. As shown in **Figure 6**d, the duty-cycled operation also demonstrated an increase in the gas response at HT with increasing temperature. To achieve sensitivity enhancement through a temperature increase of 50 °C in constant power mode, a significant power increment of 550 µW was required. Conversely, the same performance enhancement could be achieved in duty-cycling mode with a power increase of less than 30 µW. This reduction in power demand implies that, through careful power and temperature control, sensing capabilities can be improved without significantly increasing energy consumption. This also indicates the potential of this technology in developing energy-efficient environmental monitoring systems.

**Figure 6**e shows the gas responses to $NO_2$ in the RT state in duty-cycling mode, providing a comparative perspective with the HT state. Similar to the HT responses, the responses at RT also decrease with the reduction in the duty cycle. **Figure 6**f offers a direct comparison between the results obtained at HT (shown in **Figure 6**a) and RT (shown in **Figure 6**e). The responses at RT consistently outperform those at HT across the entire range of duty cycles. This is attributed to the fact that, under an intermittent RT state, the electrical resistivity of the MOS material increases instantaneously while $NO_2$ continues to adsorb on the MOS surface. Additionally, the RT state shows distinct changes in the gas response at a higher duty (20% vs. 5%) compared to the HT responses, as shown in **Figure 6**f. Concurrently, the ratio of RT and HT responses decreases with the reduction in the duty cycle (*e.g.,* 1.47 at 5% and 1.04 at 0.3% duty cycle, respectively). This is because gas desorption in the RT state lasts longer as the duty is reduced.

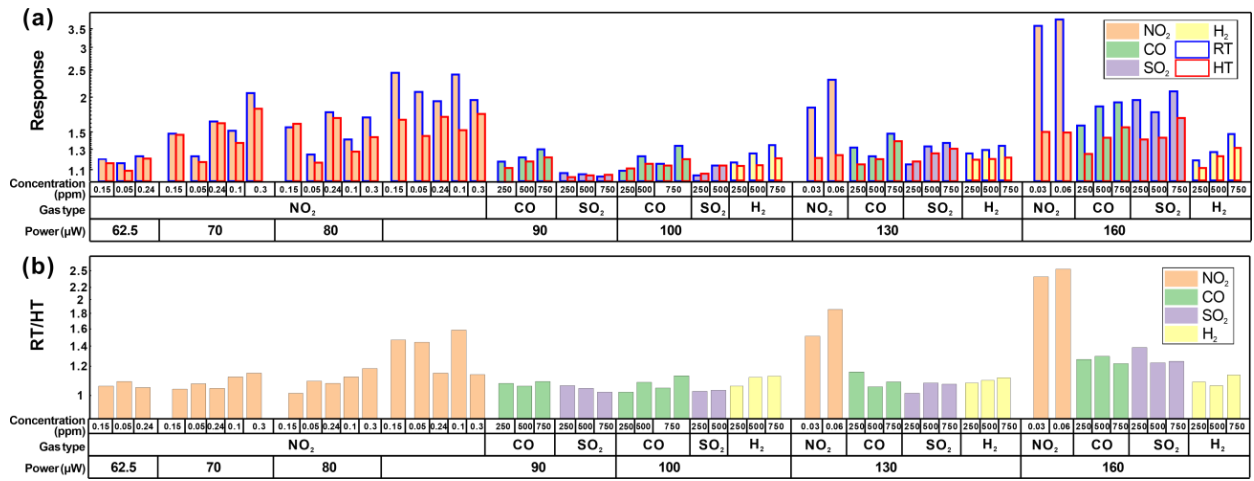

**Figure 7.** (**a**) Gas responses at HT and RT states and (**b**) their ratio (RT response/HT response) for four gases (NO$_2$, SO$_2$, CO, and H$_2$) under various gas concentrations and power conditions [body color of bars: gas species, bar outline color: power state (blue: RT, red: HT)].

**Figures 7** and **S5** show, respectively, the ratio of RT to HT responses and the transient current signals, which vary according to gas species (oxidizing gas: NO$_2$; reducing gases: SO$_2$, CO, H$_2$), their concentrations, and heater power. The charge exchange between gas analytes and MOS varies depending on the gas type. For instance, NO$_2$, an oxidizing gas, possesses a significantly higher electron affinity compared to O$_2$ (2.3 eV for NO$_2$[47,48] and 0.45 eV for O$_2$[49]), leading to direct chemisorption of the gas on the surface of the sensing material, as seen in reaction 3.[40] In other words, the reaction is more favored than those involving adsorbed reactive oxygen species (such as O$_2^-$ and O$^-$), as indicated in reactions 5 and 6.

$$NO_{2(gas)} \leftrightarrow NO_{2(ads)} \tag{2}$$

$$NO_{2(ads)} + e^-_{(CB\ of\ MOS)} \leftrightarrow NO_{2^-(ads)} \tag{3}$$

$$O_{2(gas)} \leftrightarrow O_{2(ads)} \tag{4}$$

$$O_{2(ads)} + e^-_{(CB\ of\ MOS)} \leftrightarrow O_2^-{}_{(ads)} \quad (RT\ to\ 150\ °C) \tag{5}$$

$$O_2^-{}_{(ads)} + e^-_{(CB\ of\ MOS)} \leftrightarrow 2O^-{}_{(ads)} \quad (\sim 150\ to\ \sim 500\ °C) \tag{6}$$

Conversely, reducing species such as SO$_2$, CO, and H$_2$ are oxidized through reactions with reactive oxygen species (reactions 8, 10, and 12).[50–52] Consequently, the conductivity of the MOS channel increases, as shown in **Figure S5**.

$$SO_{2(gas)} \leftrightarrow SO_{2(ads)} \tag{7}$$

$$SO_{2(ads)} + O^-{}_{(ads)} \leftrightarrow SO_{3(ads)} + e^- \tag{8}$$

$$CO_{(gas)} \leftrightarrow CO_{(ads)} \tag{9}$$

$$CO_{(ads)} + O^-_{(ads)} \leftrightarrow CO_{2(ads)} + e^- \tag{10}$$

$$H_{2(gas)} \leftrightarrow H_{2(ads)} \tag{11}$$

$$H_{2(ads)} + O^-_{(ads)} \leftrightarrow H_2O_{(ads)} + e^- \tag{12}$$

Therefore, the patterns of gas adsorption/desorption and electron transport can vary depending on the gas type, influenced by the disparity in electron affinity values between gases (2.3 eV for $NO_2$, 1.1 eV for $SO_2$,[53–55] 1.3 eV for CO,[56] 2.0 eV for $H_2$,[57] and 0.45 eV for $O_2$). This results in subtle differences in sensing kinetics. Additionally, the rate of adsorption and desorption, as well as the activation energy required for these reactions, can all be influenced by temperature changes.[19,40] Thus, the ratio of RT to HT responses varies according to gas species, concentration, and heater temperature. These unique traits suggest the potential of analyzing duty-cycling-driven gas responses as a novel strategy for enhancing gas selectivity.

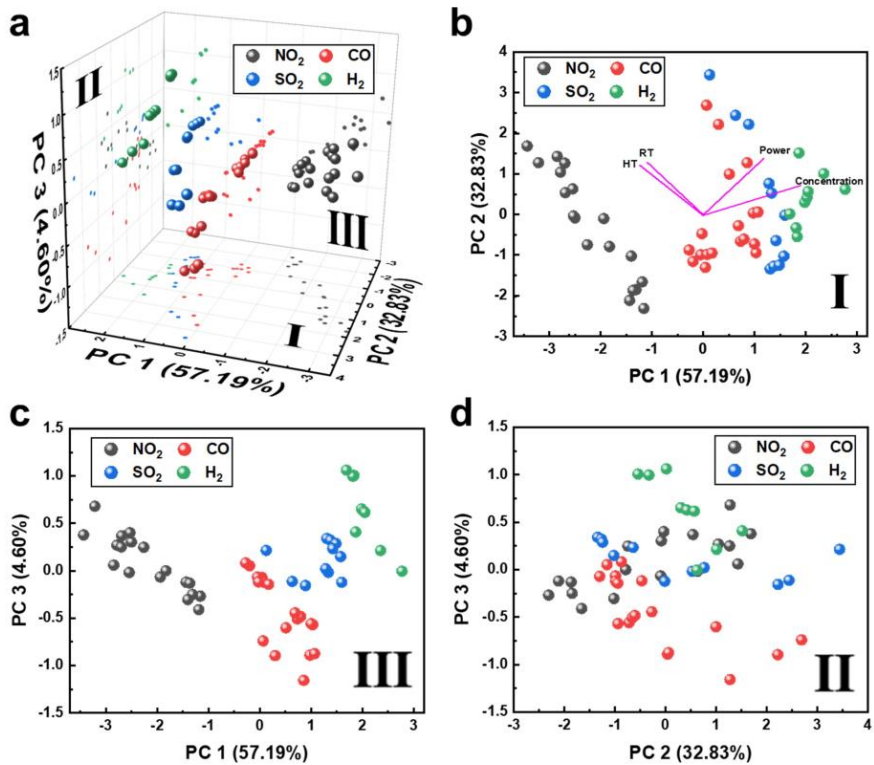

**Figure 8**. Principal component analysis (PCA) plots of gas responses for four gases, including NO$_2$, SO$_2$, CO, and H$_2$ (PC1: first principal component, PC2: second principal component, PC3: third principal component): (**a**) 3D pot of PCA; (**b**–**d**) corresponding 2D projections on the (**b**) PC1 × PC2, (**c**) PC1 × PC3, and (**d**) PC2 × PC3 planes. PCA was performed using the complete set of variables, including responses at HT and RT, power, and concentration.

Building upon this, the gas categorization capability of duty-cycling-driven sensor operation was evaluated using 3D plots of gas responses at RT and HT and the response ratio according to gas types and their concentrations, as shown in **Figure S6**. Despite the distinct response behaviors among gas species, these plots were not sufficiently characteristic to identify gas species and concentrations through simple correlation. Instead, the single sensor-driven gas responses were analyzed using principal component analysis (PCA), resulting in the primary components of PC1 (57.19%), PC2 (32.83%), and PC3 (4.60%). In the PCA plots, shown in **Figure 8**, data points are distributed according to gas categories. However, identification of gas types and concentrations remains challenging with merely PCA analysis results. In addition, this method does not meet the need for real-time gas identification, as it requires waiting until the gas reaction has completely saturated. To overcome this, we employed a CNN, a deep learning model for image recognition and pattern analysis, which will be described in the following section.

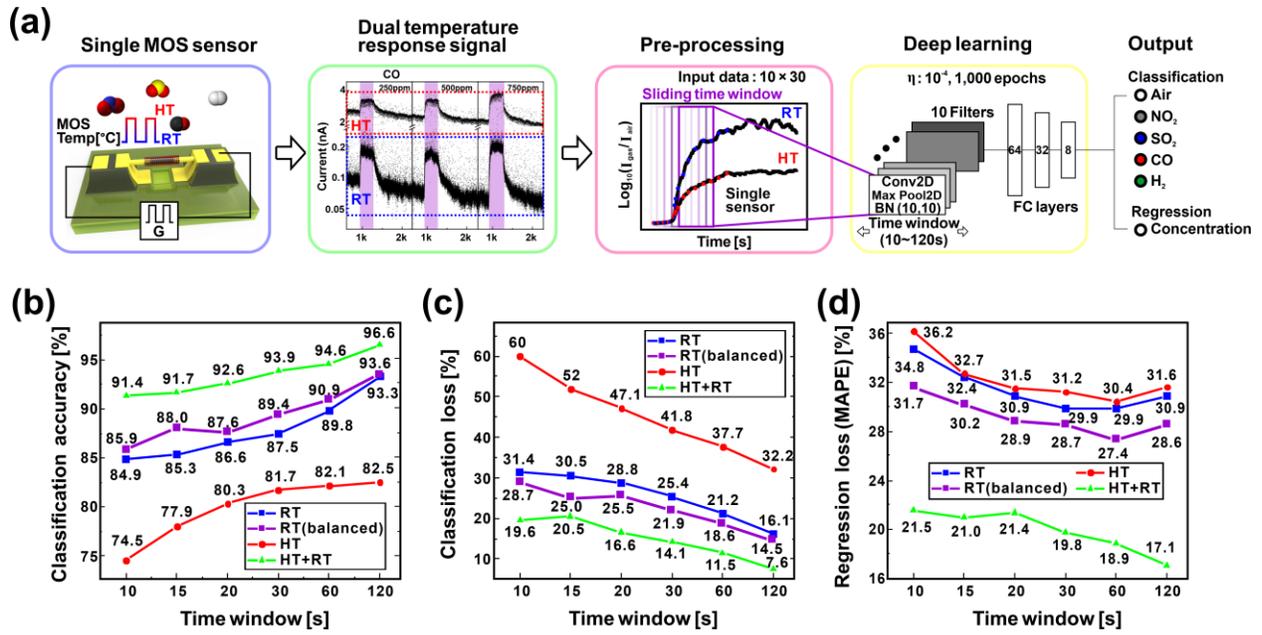

**Figure 9**. Single sensor-based gas identification for five gases (air, $NO_2$, $SO_2$, CO, $H_2$) through a CNN algorithm using a duty-cycling-driven gas sensor (duty cycle = 10%, duty frequency = 1 Hz, power = $NO_2$: 90 μW; $SO_2$, CO, $H_2$: 160 μW): (**a**) Conceptual architecture of the CNN-based gas classification and concentration regression. Comparisons of (**b**) gas type classification accuracy, (**c**) classification loss, and (**d**) regression loss results between the RT response (blue), HT response (red), a single combined response (HT+RT response; green), and balanced RT response, the data amount of which matches that of the HT+RT response.

**CNN-based gas identification using dual responses derived from a single sensor.** As mentioned earlier, the distinctive dual responses derived from a duty-cycling-driven MOS gas sensor have shown considerable promise in the accurate identification of gas types and concentrations through ML. In addition to PCA, various ML algorithms, such as the K-nearest neighbor (KNN)[58] and the support vector machine (SVM)[59], have been widely utilized for gas identification. However, these methods have limitations in their application to real-time gas identification and concentration prediction. This is because they use the gas responses measured after the gas reactions have been completed as training data, which requires waiting until the gas response has reached saturation. In addition, for the same reason, gas sensors with relatively long response times are unsuitable for these methods. Instead, the spiking neural network (SNN)[60] and CNN[61–63] have been actively studied as real-time gas identification tools. Among them, the CNN algorithm converts the gas responses into 2D spectrogram images and analyzes them in real time by identifying the gas within the sliding time window. This method is renowned for robust image recognition and pattern analysis, as explained in detail in **Supplementary Note 1**. Recently, studies on CNN-based real-time gas identification using a micro-LED-integrated MOS sensor array were reported.[61,62] Owing to the energy-efficient photoactivation by the LED, this approach achieved significant power savings despite utilizing a multi-sensor array for accurate gas prediction. Nonetheless, sensor arrays necessitate extra costs for the sensor system's fabrication, packaging, and operation. The same research group also reported a single-sensor-based e-nose system, enabling gas identification through the application of a fast-changing pseudorandom voltage input to the LED.[63] Despite using a single sensor, the power consumption was greater than that of multi-sensor systems because the fast time-variant illumination increased the total energy usage.

In this study, gas classification and concentration regression were conducted by analyzing the duty-cycling-derived dual responses of a single MOS sensor using the CNN algorithm. Additionally, the duty-cycling mode (duty = 10%, power = 62.5–160 μW) offers the crucial advantage of reducing power consumption to a level comparable to that of the LED-driven MOS sensor array. **Figure 9**a shows a schematic of the CNN-based gas analysis process. First, the gas sensing data collected from a single sensor was preprocessed before training the CNN model to enhance learning efficiency and accuracy. This involved data collection balancing the amount of acquired data between gases such as air, $SO_2$, CO, $H_2$, and $NO_2$ to ensure fair learning of each gas. Additionally, the training gas response data were normalized to $Log_{10}$ (gas response) [*e.g.*, $Log_{10}(I_a/I_g)$ for oxidizing gases and $Log_{10}(I_g/I_a)$ for reducing gases] to prevent the overestimation of the signal of highly responsive gases, such as $NO_2$. Subsequently, the normalized response data were captured by sliding a time window and converted into an input matrix for the training. Data were acquired within a unit time window (*e.g.*, 10, 15, 20, 30, 60, or 120 s) and were renewed every 1 s (stride interval = 1 s). The matrix row was filled with data representing 10 samples per second. This sampling rate (10 Hz) was set to match the duty-cycling frequency for real-time monitoring. The composition of the matrix was determined according to the duty cycles and sliding time windows. For example, in the case of 10% duty cycle and a time window of 30 s with a sampling rate of 10 Hz, the row is composed of 1 HT and 9 RT data points because the ratio of the HT data and RT data is 1:9, and the number of columns is set to 30 (s), resulting in a 10 × 30 matrix. The time window size was optimized by evaluating the classification and regression accuracies according to the time window size, as described in the following paragraph. For training and validation, 70% and 30% of the time series input matrices were used, respectively. For gas classification training, air, $SO_2$, CO, $H_2$, and $NO_2$ were labeled 0, 1, 2, 3, and 4, respectively. For the gas concentration regression, the concentrations of each gas type were normalized using min-max normalization, adjusting the concentration range to a uniform scale. Subsequently, these adjusted concentration levels were used as concentration labels, such that the concentration values across different gas types were comparable and standardized, facilitating a more accurate regression analysis. After the pre-processing steps, the input data were processed by the convolution kernels and average pooling layers to retain prominent features and reduce data volume. The first layer of the CNN used ten 10 × 10 kernels, and the second layer utilized ten 1 × 3 kernels. For chemical reactions that vary over time, convolution layers are more suitable for extracting temporal characteristics.[64] The output data processed through these convolutional layers were then amalgamated and forwarded to fully connected (FC) layers of 32, 16, and 8 nodes. Each hidden layer of the CNN and FC layers employed batch normalization (BN) and the Leaky-ReLU activation function to improve the model's learning stability and performance. The output layer of our model was configured to consist of six nodes: five for classifying the gas types and an additional one for determining the gas concentration. This setup allowed us to observe and analyze the model's effectiveness in both gas type classification and concentration determination. The classification of gas types was facilitated by the softmax function, which identifies the gas type with the highest probability of presence. A cross-entropy loss function was utilized to effectively measure the difference between the model's predicted probability distribution and the actual labels. The regression node used the MAPE loss function to evaluate the accuracy of the regression model's predictions. This loss function measures the average absolute difference between the predicted and actual gas concentrations. Model training was executed using an Adam optimizer (initial learning rate $\eta = 0.0001$) while adjusting the learning rate and managing sparse gradients with the goal of minimizing the total loss. The CNN model evaluation factors, such as classification accuracy, classification loss (cross-entropy loss), and regression loss (MAPE loss), converged at an epoch of 1000 without a significant divergence between training and validation data, as shown in **Figure S7**. This indicates that the CNN model was well-constructed (with, *e.g.*, a balanced loss weighting and the Adam optimizer set at a low learning rate) and effectively generated such that it adapted and responded accurately to new and unseen data.[65] The gas classification and concentration prediction were determined simultaneously through the forward propagation of one cycle of gas test data set into the pre-trained CNN model. In the regression, the gas concentration was set to 0 if the gas type was identified as air. The regression efficiency (measured as MAPE loss) was calculated only for data points where gas responses were detected, excluding air data.

The efficiency of gas identification based on dual-response signals (denoted as HT+RT signals) derived from a single sensor was assessed by comparing it with identification results obtained from either the HT signal or the RT signal. It is important to note that due to the duty-cycling operations, the amount of data differed among RT, HT, and HT+RT data. For instance, for a duty cycle of 10%, the data ratios for HT, RT, and HT+RT are 1:9:10. Thus, to ensure a fair comparison of the CNN results from HT+RT and RT data, the amount of data for the RT state was adjusted to match that of the HT+RT data, and this adjusted data was denoted as "balanced RT." However, at HT, the amount of data is limited for small duties, so HT data was not adjusted. All datasets exhibit enhancements in gas identification in terms of classification accuracy, classification loss, and regression loss with increases in the size of the time window, as shown in **Figures 9**b–d. Notably, RT data showed a significant enhancement in gas identification over HT data because the amount of the former is nine times greater than that of the latter. More importantly, the dual-response signal (HT+RT data) showed better gas identification capabilities compared to the other datasets, even with a slight enhancement of the balanced RT results over the RT data results. The CNN model based on HT+RT data falsely predicted $SO_2$ and CO, whereas it could accurately distinguish air, $H_2$, and $NO_2$, as shown in the confusion matrices of **Figure S8**. This is due to the similarity in electron affinities of those two gases ($SO_2$: 1.1 eV, CO: 1.3 eV).[53–56] In particular, the dual-response signal (HT+RT) showed superior concentration prediction capability compared to other datasets. This excellence in gas identification using the HT+RT data is attributed to the unique features, such as differences in response patterns (*e.g.,* response magnitude, response speed, noise) between the HT and RT data according to gas type and concentration, as indicated in **Figures S5 and 7**. As shown in **Figures 9**b–d and **S8**, gas classification and concentration prediction accuracies deteriorate as the time window decreases. Thus, a reduced time window size diminishes the overall prediction time but yields lower accuracy. Conversely, a larger time window size enhances accuracy by conducting supervised learning with a richer transient sensor signal dataset, yet it escalates the risk of overfitting and extends the prediction time. Therefore, the size of the sliding time window should be determined based on the target applications. In situations where rapid gas prediction is required, such as for toxic or explosive gases, a smaller time window is recommended. Conversely, if accurate gas identification is required, a larger time window can be used. When HT and balanced RT were used separately, the classification accuracies were only 81.7% and 89.4%, respectively, with a time window of 30 s. In contrast, the HT+RT model achieved gas classification accuracy over 93% with a 14.1% loss in addition to less than 20% regression loss, highlighting its effectiveness, especially for early diagnosis scenarios where shorter time windows are crucial.

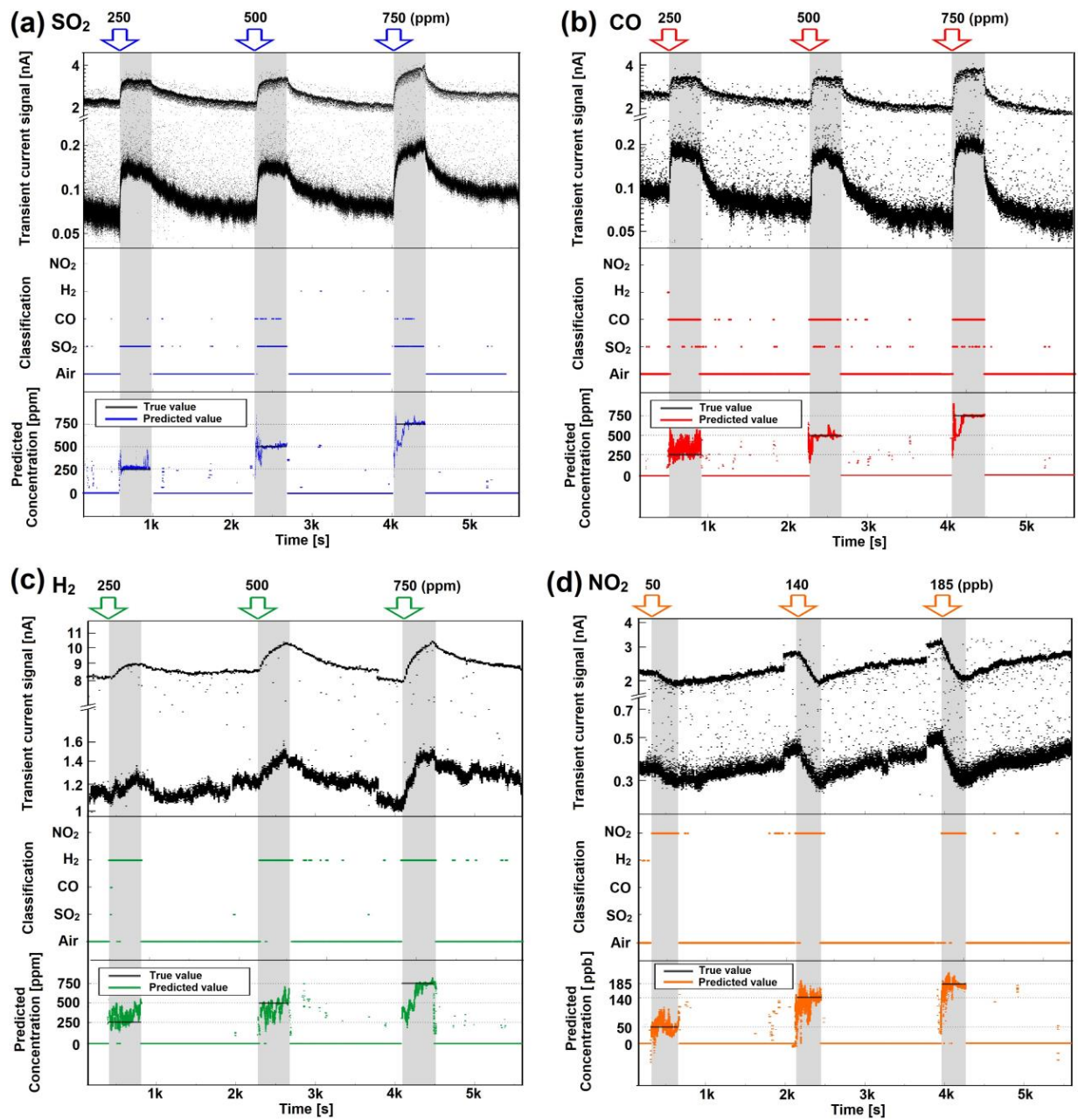

**Figure 10.** Real-time prediction of gas type and concentration for five gases [air, (a) SO$_2$, (b) CO, (c) H$_2$, (d) NO$_2$] using dual-response signals (time window = 30 s, duty cycle = 10%, duty frequency = 1 Hz).

**Figure 10** shows the real-time prediction of gas types and the corresponding concentrations for five gases (air, $SO_2$, CO, $H_2$, and $NO_2$) using dual-response signals. In the classification and regression graphs, the true value is depicted by a black line, whereas a colored dot represents the CNN prediction results. The CNN model exhibited the capability for real-time prediction of gas type and concentration, even though the gas response required extended time for signal saturation with changes in gas conditions, as shown in the transient current signal graphs in **Figure 10**. This highlights that CNN-based gas prediction, in conjunction with a duty-cycling-driven single sensor, can effectively address the slow response and recovery times typically associated with MOS-based gas sensors.

**Table 1** compares the proposed single-sensor-based e-nose system with recently reported ML-based e-nose systems.[58–63] These studies employed miniaturized heaters or micro-LEDs to save power while ensuring accurate prediction mainly through multi-sensor operation. Regarding power consumption, micro-LED technology significantly outperformed microheaters, being approximately 100 times more efficient. Additionally, the micro-LED-based approaches achieved gas identification using a considerably smaller number of sensors. Furthermore, SNN and CNN have been actively utilized for real-time identification while achieving high accuracy. In this study, we demonstrated an ultralow power e-nose system by duty cycling a single nanoheater-embedded MOS sensor whose power efficiency is comparable to that of state-of-the-art micro-LED-based systems. Moreover, the distinctive signals generated by duty cycling, combined with CNN-based ML techniques, enable real-time gas identification using just a single sensor. Additionally, this single-sensor operation allows for cost-effective implementation of e-nose systems, in addition to saving power.

**Table 1.** Comparison of the proposed system with other machine-learning-based e-nose systems.

| | Li et al.[58] | Thai et al.[59] | Kwon et al.[60] | Kang et al.[61] | Lee et al.[62] | Cho et al.[63] | **This work** |
|---|---|---|---|---|---|---|---|
| Sensor type | Microheater +MOS | Microheater +MOS | Microheater +MOS | Microheater +MOS | Micro-LED +MOS | Micro-LED +MOS | Nanoheater + MOS |
| # of sensors | 2 | 8 | 12 | 8 | 2 | 1 | 1 |
| *# of target gases | 4 | 6 | 3 | 6 | 5 | 4 | 5 |
| Total sensor power | 34 mW | 340 mW | ~12 mW | 88 mW | 380 µW | 526 µW | 160 µW (duty cycle = 10%) |
| Analysis method | KNN | SVM | SNN | CNN | CNN | D-CNN | CNN |
| Classification accuracy [%] | 99.86 | 100 | - | 98.1 | 99.32 | 96.53 | 93.9 |
| Regression error (MAPE, %) | 2.08–5.44 | 8–28 | 3 | 10.15 | 13.82 | 31.99 | 19.8 |
| Real-time prediction | X | X | O | O | O | O | O |

[a]MOS: metal oxide semiconductor; KNN: k-nearest neighbor; SVM: support vector machine; SNN, spiking neural network; CNN: convolutional neural network; D-CNN: deep convolutional neural network; MAE: mean absolute error; MAPE: mean absolute percentage error. *Air is included in the target gases.

METHODS

**Simulation of the Temperature Profile of the Sensor.** The temperature profile of the suspended 1D nanoheater coated with a MOS NW network was simulated using a commercial software (COMSOL Multiphysics 5.4, COMSOL, Stockholm, Sweden). In this study, the heat transfer and electric current modules in the software were coupled for the simulation. Convection and radiation effects were ignored owing to the relatively low heater temperature and the low Rayleigh number resulting from the small heater size, respectively. The simulated model consisted of a suspended carbon nanowire backbone (width = 220 nm, thickness = 350 nm, length = 175 μm, separation from the substrate = 6 μm), a 50 nm-thick Au heater layer, a 50 nm-thick $HfO_2$ insulation layer, 100 nm-thick Au sensor electrode leads, and a ZnO NW network [simplified as a hollow cylinder (outer diameter = 1.4 μm)]. The simulation was performed under atmospheric conditions of 1 atm air and 293 K.

**Fabrication of the 1D Nanoheater-Embedded MOS Gas Sensor Platform.** As illustrated in **Figure S1,** the sensor platform was manufactured through three main wafer-level nanofabrication steps: forming a suspended carbon nanowire backbone, integrating a nanoheater layer, and growing an MOS NW network. All the patterning processes were performed using a conventional microscale mask aligner (MA/BA6-8, SUSS MicroTec. Inc., Germany). Initially, a built-in shadow mask was fabricated to facilitate the selective heater layer coating on it (**Figure S1**a). A 1 μm-thick $SiO_2$ insulation layer was grown on a 6-inch Si wafer using wet oxidation. Subsequently, a $SiO_2$ etch mask was patterned through buffered oxide etching (BOE, Baker Chemical Co., Ltd., USA) under a positive photoresist (AZ-5214E, AZ Electronic Materials, USA) mask. The Si substrate was then etched in an isotropic manner under a $SiO_2$ mask using deep reactive ion etching (Tegal 200 SE DRIE, Tegal Corp., USA), forming a built-in shadow mask comprising $SiO_2$ eaves and Si undercut structures (**Figure S2**f inset). Above this built-in shadow mask, a suspended 1D carbon backbone was fabricated using C-MEMS, involving two successive photolithography steps and a single pyrolysis process (**Figure S1**b). A 23 μm-thick negative photoresist (SU-8 2025, MicroChem Corp., USA) was spin-coated onto the substrate and soft-baked. This negative photoresist layer was UV-exposed (250 mJ cm$^{-2}$) entirely from top to bottom to produce polymer post structures. A subsequent second UV exposure with a lower dose (10 mJ cm$^{-2}$) was performed between the pre-exposed post structures to form a suspended micrometer-sized polymer wire. After post-exposure baking, a monolithic polymer structure consisting of a micrometer-sized suspended polymer wire and polymer posts was formed through a single development step. This monolithic polymer structure was then carbonized by pyrolysis at 600 °C in a vacuum furnace. A gold heater line was selectively coated on the suspended carbon backbone by photolithography (NR9-8000) and e-beam evaporation (10 nm Cr/50 nm Au), as shown in **Figure S1**c. After removing the negative photoresist layer, a 50 nm-thick HfO$_2$ insulation layer was deposited on the entire substrate *via* atomic layer deposition (Lucida D100, NCD Co., Ltd., Republic of Korea). The HfO$_2$ layer was patterned using photolithography and reactive ion etching processes for electrical connection to the nanoheater (**Figure S1**d). Sensor electrode leads (20 nm Cr/200 nm Au) for connection to the sensing material (ZnO NWs) were patterned similarly to the patterning process of the heater electrodes (**Figure S1**e). Finally, ZnO NWs were regioselectively grown in the central 60 μm of the nanoheater through selective seed layer deposition and hydrothermal growth (**Figure S1**f). A ZnO seed layer (thickness = 20 nm) was deposited *via* RF sputtering (SRN-120, Sorona Co., Ltd., Republic of Korea), and ZnO NWs were grown using zinc nitrate hexahydrate (Zn(NO$_3$)$_2$·6H$_2$O, Sigma-Aldrich, USA) and hexamethylenetetramine (HMTA, (CH$_2$)$_6$N$_4$, Sigma-Aldrich, USA) in deionized water in an autoclave system. For the selective seed layer deposition, a positive photoresist (AZ 9260) mask was patterned before the seed layer deposition, ensuring that the photoresist remained below the suspended nanoheater by controlling the UV exposure energy.

**Characterization of Morphology and Material Composition.** The morphology of the sensor was characterized using SEM (Quanta 200, FEI, USA). The chemical composition of the present sensor was analyzed using EDS (Silicon Drift Detect–X-MAXN, Oxford Instruments plc, UK). XPS (K-alpha, ThermoFisher Scientific Corp., USA) analysis was performed to determine the chemical states and compositions of the ZnO NWs and pyrolyzed carbon.

**Experimental Setup for the Gas Sensor Test.** In the gas sensor test, the target gas concentration was controlled by mixing the gas with compressed dry air using mass flow controllers (GMC1200, Atovac, Republic of Korea), as shown in **Figure S9**. The ZnO NW network was heated by applying a square current wave (Agilent 33220A, Agilent Technologies, Inc., USA) to the 1D nanoheater. Simultaneously, the change in the ZnO NW network resistance was monitored using a source meter (Keithley 2450, Keithley Instruments, Inc., USA). The thermal time constant of the nanoheater was evaluated in dry air at RT using a pulsed current signal (0.1–4 mA, 0.001–1000 Hz) generated from a current source (Keithley 6221 AC and DC source, Keithley Instruments, Inc., USA). The heating/cooling rate of the ZnO NWs was measured by recording the voltage output every 10 ns with an oscilloscope (DSO7032A InfiniiVision, Agilent Technologies, Inc., USA), while a constant current (100 nA) from a current source (Keithley 2450, Keithley Instruments, Inc., USA) flowed through the ZnO NWs.

**Deep-learning-based Gas Prediction**. A CNN model was constructed using Python and open-source libraries. TensorFlow (Google, USA) was used as the primary ML framework. The training of the CNN model and the analysis of the gas sensing results were conducted in a high-performance computing environment, utilizing a GeForce RTX 3050 GPU (NVIDIA, USA).

CONCLUSION

This study successfully addressed the challenges of high power consumption and limited selectivity of conventional MOS gas sensors by implementing a nanoheater-embedded gas sensor driven by duty cycling. Our approach, involving the supply of repeated pulsed-power inputs to the suspended 1D nanoheater, effectively decoupled the effects of temperature and surface charge exchange on the sensor's response. This was due to the nanoheater's thermal response being significantly faster than redox kinetics, providing a dual-response sensing signal within a single time domain. The distinctive features extracted from the dual responses, including the ratio of data at power-off and power-on states according to gas types and concentrations, remained distinguishable even after signal normalization for deep learning efficiency and stability. This enabled the real-time identification of five gas types (air, $NO_2$, $SO_2$, CO, and $H_2$) and their concentrations through a CNN. Moreover, the combination of ultra-small size and ultra-fast switching capability enabled the sensor to operate with extremely low power consumption (160 μW) under duty-cycling mode without significantly compromising sensor performance. This level of power efficiency surpasses that of state-of-the-art e-nose systems. In addition, the MOS nanomaterial-based sensors built on a suspended 1D nanoheater were manufactured using wafer-level batch microfabrication processes despite their complex 3D mixed-scale architecture. These advantages of the duty-cycling-driven, single-sensor-based e-nose system offer a promising solution for a cost-effective implementation of real-time gas monitoring systems capable of operating at ultralow power, which are well-suited for battery-driven mobile and IoT systems.

ASSOCIATED CONTENT

Data Availability Statement

All methods and data supporting the findings of this study are available within the paper and its Supporting Information. All raw data and code are available upon request.

**Supporting Information Available**: <Convolution neural networks (CNN) for image processing; schematic of the fabrication steps of a metal-oxide nanowire-based gas sensor with an embedded nanoheater; scanning electron microscopy images of a suspended metal-oxide nanowire-based gas sensor with an embedded nanoheater; schematic of selective metal coating on a 1D carbon nanostructure using a built-in shadow mask and an evaporation method; diagrams illustrating the dependency of the sensor signals on the number of readout points in duty-cycling mode; transient dual-response current signals measured at various concentrations of reducing gases; comparison of gas ($NO_2$, CO, $SO_2$, and $H_2$) responses at HT, RT, and their ratio (RT/HT) with respect to power consumption and concentration; evaluation of the effectiveness of gas identification using dual-response signals from a single MOS sensor through a CNN algorithm; confusion matrices of gas type identification for two time windows; gas sensing test setup; comparison of the proposed and state-of-the-art duty-cycled MOS gas sensors (Table S1)> (PDF)

AUTHOR INFORMATION

Corresponding Author

*Heungjoo Shin – *Department of Mechanical Engineering, Ulsan National Institute of Science and Technology (UNIST), Ulsan, 44919, Republic of Korea*; orcid.org/0000-0002-5660-7093; Email: hjshin@unist.ac.kr

*Jae Joon Kim – *Department of Electrical Engineering, Ulsan National Institute of Science and Technology (UNIST), Ulsan, 44919, Republic of Korea*; orcid.org/0000-0003-4581-4115; Email: jaejoon@unist.ac.kr

Authors

Taejung Kim – *Department of Mechanical Engineering, Ulsan National Institute of Science and Technology (UNIST), Ulsan, 44919, Republic of Korea*; orcid.org/0000-0002-2433-7295

Yonggi Kim – *Department of Electrical Engineering, Ulsan National Institute of Science and Technology (UNIST), Ulsan, 44919, Republic of Korea*

Wootaek Cho – *Department of Mechanical Engineering, Ulsan National Institute of Science and Technology (UNIST), Ulsan, 44919, Republic of Korea*

Jong-Hyun Kwak – *Department of Mechanical Engineering, Ulsan National Institute of Science and Technology (UNIST), Ulsan, 44919, Republic of Korea*

Jeonghoon Cho – *Department of Electrical Engineering, Ulsan National Institute of Science and Technology (UNIST), Ulsan, 44919, Republic of Korea*

Youjang Pyeon – *Department of Electrical Engineering, Ulsan National Institute of Science and Technology (UNIST), Ulsan, 44919, Republic of Korea*

Author Contributions

‡ T. Kim and Y. Kim contributed equally to the work. The manuscript was written through contributions of all authors. T. Kim conducted all experiments, analyzed the data, and wrote the paper. Y. Kim, J. Cho, and Y. Pyeon discussed the development of the machine learning algorithm. W. Cho performed simulation of temperature profile of the sensor. T. Kim, J.-H. Kwak, and W. Cho discussed the fabrication of the gas sensors. J. J. Kim and H. Shin supervised the project. All authors have given approval to the final version of the manuscript.

Funding Sources

This research was supported by Basic Science Research Program through the National Research Foundation of Korea (NRF), funded by the Ministry of Education (2020R1A6A1A03040570), and the Technology Innovation Program (00144157, Development of Heterogeneous Multi-Sensor Micro-System Platform) funded by the Ministry of Trade, Industry and Energy, Republic of Korea.

Notes

The authors declare no competing financial interest.

ACKNOWLEDGMENT

This research was supported by Basic Science Research Program through the National Research Foundation of Korea (NRF), funded by the Ministry of Education (2020R1A6A1A03040570), and the Technology Innovation Program (00144157, Development of Heterogeneous Multi-Sensor Micro-System Platform) funded by the Ministry of Trade, Industry and Energy, Republic of Korea. We are grateful for the technical assistance received from the staff members at UCRF (UNIST Central Research Facilities) in UNIST.

REFERENCES

[1] Lewis, A.; Edwards, P. Validate Personal Air-Pollution Sensors. *Nature* **2016**, 535, 29–31. DOI: 10.1038/535029a

[2] van den Broek, J.; Abregg, S.; Pratsinis, S. E.; Guntner, A. T. Highly Selective Detection of Methanol over Ethanol by a Handheld Gas Sensor. *Nat. Commun.* **2019**, *10*, 4220. DOI: 10.1038/s41467-019-12223-4

[3] Kim, H. J; Lee, J. H. Highly Sensitive and Selective Gas Sensors Using P-Type Oxide Semiconductors: Overview. *Sens. Actuators, B* **2014**, *192*, 607–627. DOI: 10.1016/j.snb.2013.11.005

[4] Tao, N. Challenges and Promises of Metal Oxide Nanosensors. *ACS Sens.* **2019**, *4*, 780–780. DOI: 10.1021/acssensors.9b00622

[5] Chen, J.; Chen, Z.; Boussaid, F.; Zhang, D.; Pan, X.; Zhao, H.; Bermak, A.; Tsui, C. -Y.; Wang, X.; Fan, Z. Ultra-Low-Power Smart Electronic Nose System Based on Three-Dimensional Tin Oxide Nanotube Arrays. *ACS Nano* **2018**, *12*, 6079–6088. DOI: 10.1021/acsnano.8b02371

[6] Ponzoni, A.; Baratto, C.; Cattabiani, N.; Falasconi, M.; Galstyan, V.; Nunez-Carmona, E.; Rigoni, F.; Sberveglieri, V.; Zambotti, G.; Zappa, D. Metal Oxide Gas Sensors, a Survey of Selectivity Issues Addressed at the SENSOR Lab, Brescia (Italy). *Sensors* **2017**, *17*, 714. DOI: 10.3390/s17040714

[7] Kumar, R.; Liu, X.; Zhang, J.; Kumar, M. Room-Temperature Gas Sensors Under Photoactivation: From Metal Oxides to 2D Materials. *Nano-Micro Lett.* **2020**, *12*, 1–37. DOI: 10.1007/s40820-020-00503-4

[8] Wang, Z.; Bu, M.; Hu, N.; Zhao, L. An Overview on Room-Temperature Chemiresistor Gas Sensors Based on 2D Materials: Research Status and Challenge. *Composites, Part B* **2023**, *248*, 110378. DOI: 10.1016/j.compositesb.2022.110378

[9] Fàbrega, C.; Casals, O.; Hernández-Ramírez, F.; Prades, J. D. A Review on Efficient Self-Heating in Nanowire Sensors: Prospects for Very-Low Power Devices. *Sens. Actuators, B* **2018**, *256*, 797–811. DOI: 10.1016/j.snb.2017.10.003

[10] Bhattacharyya, P. Technological Journey Towards Reliable Microheater Development for MEMS Gas Sensors: A Review. *IEEE Trans. Device Mater. Reliab.* **2014**, *14*(2), 589–599. DOI: 10.1109/TDMR.2014.2311801.

[11] Dey, A. Semiconductor Metal Oxide Gas Sensors: A Review. *Mater. Sci. Eng., B* **2018**, *229*, 206−217. DOI: 10.1016/j.mseb.2017.12.036

[12] Long, H.; Turner, S.; Yan, A.; Xu, H.; Jang, M.; Carraro, C.; Maboudian, R.; Zettl, A. Plasma Assisted Formation of 3D Highly Porous Nanostructured Metal Oxide Network on Microheater Platform for Low Power Gas Sensing. *Sens. Actuators, B* **2019**, *301*, 127067. DOI: 10.1016/j.snb.2019.127067

[13] Xie, D.; Chen, D.; Peng, S.; Yang, Y.; Xu, L.; Wu, F. A Low Power Cantilever-Based Metal Oxide Semiconductor Gas Sensor. *IEEE Electron Device Lett.* **2019**, *40*, 1178–1181. DOI: 10.1109/LED.2019.2914271

[14] Choi, K. W.; Jo, M. S.; Lee, J. S.; Yoo, J. Y.; Yoon, J. B. Perfectly Aligned, Air-Suspended Nanowire Array Heater and Its Application In an Always-On Gas Sensor. *Adv. Funct. Mater.* **2020**, *30*(39), 2004448. DOI: 10.1002/adfm.202004448

[15] Long, H.; Harley-Trochimczyk, A.; He, T.; Pham, T.; Tang, Z.; Shi, T.; Zettl, A.; Mickeslon, W.; Carraro, C.; Maboudian, R. In Situ Localized Growth of Porous Tin Oxide Films on Low Power Microheater Platform for Low Temperature CO Detection. *ACS Sens.* **2016**, *1*(4), 339–343. DOI: 10.1021/acssensors.5b00302

[16] Zhou, Q.; Sussman, A.; Chang, J.; Dong, J.; Zettl, A.; Mickelson, W. Fast Response Integrated MEMS Microheaters for Ultra Low Power Gas Detection. *Sens. Actuators, A* **2015**, *223*, 67–75. DOI: 10.1016/j.sna.2014.12.005

[17] Suematsu, K.; Harano, W.; Oyama, T.; Shin, Y.; Watanabe, K.; Shimanoe, K. Pulse-Driven Semiconductor Gas Sensors Toward PPT Level Toluene Detection. *Anal. Chem.* **2018**, *90*(19), 11219–11223. DOI: 10.1021/acs.analchem.8b03076

[18] Kalantar-Zadeh, K.; Berean, K. J.; Ha, N.; Chrimes, A. F.; Xu, K.; Grando, D.; Ou, J. Z.; Pillai, N.; Campbell, J. L.; Brkljača, R.; Taylor, K. M.; Burgell, R. E.; Yao, C. K.; Ward, S. A.; McSweeney, C. S.; Muir, J. G.; Gibson, P. R. A Human Pilot Trial of Ingestible Electronic Capsules Capable of Sensing Different Gases in the Gut. *Nature Electronics* **2018**, *1*(1), 79–87. DOI: 10.1038/s41928-017-0004-x

[19] Tang, W.; Chen, Z.; Song, Z.; Wang, C.; Wan, Z. A.; Chan, C. L. J.; Chen, Z.; Ye, W.; Fan, Z. Microheater Integrated Nanotube Array Gas Sensor for Parts-Per-Trillion Level Gas Detection and Single Sensor-Based Gas Discrimination. *ACS Nano* **2022**, *16*(7), 10968–10978. DOI: 10.1021/acsnano.2c03372

[20] Shin, W.; Jung, G.; Hong, S.; Jeong, Y.; Park, J.; Kim, D.; Park, B. -G.; Lee, J. H. Optimization of Channel Structure and Bias Condition for Signal-To-Noise Ratio Improvement in Si-Based FET-Type Gas Sensor with Horizontal Floating-Gate. *Sens. Actuators, B* **2022**, *357*, 131398. DOI: 10.1016/j.snb.2022.131398

[21] Kim, S. H.; Jo, M. S.; Choi, K. W.; Yoo, J. Y.; Kim, B. J.; Yang, J. S.; Chung, M. K.; Kim, T. S.; Yoon, J. B. Ultrathin Serpentine Insulation Layer Architecture for Ultralow Power Gas Sensor. *Small* **2023**, 2304555. DOI: 10.1002/smll.202304555

[22] Persaud, K.; Dodd, G. Analysis of Discrimination Mechanisms in the Mammalian Olfactory System Using a Model Nose. *Nature* **1982**, *299*, 352–355. DOI: 10.1038/299352a0

[23] Zhou, K.; Liu, Y. Early-Stage Gas Identification Using Convolutional Long Short-Term Neural Network with Sensor Array Time Series Data. *Sensors* **2021**, *21*, 4826. DOI: 10.3390/s21144826

[24] Kwon, Y. M.; Oh, B.; Purbia, R.; Chae, H. Y.; Han, G. H.; Kim, S. W.; Choi, K. -J.; Lee, Y.; Kim, J. J.; Baik, J. M. High-Performance and Self-Calibrating Multi-Gas Sensor Interface to Trace Multiple Gas Species with Sub-PPM Level. *Sens. Actuators, B* **2023**, *375,* 132939. DOI: 10.1016/j.snb.2022.132939

[25] Sears, W. M.; Colbow, K.; Consadori, F. Algorithms to Improve the Selectivity of Thermally-Cycled Tin Oxide Gas Sensors. *Sens. Actuators* **1989**, *19*, 333–349. DOI: 10.1016/0250-6874(89)87084-2

[26] Lee, A. P.; Reedy, B. J. Temperature Modulation in Semiconductor Gas Sensing. *Sens. Actuators, B* **1999***, 60*(1), 35–42. DOI: 10.1016/S0925-4005(99)00241-5

[27] Chakraborty, S.; Sen, A; Maiti, H. S. Selective Detection of Methane and Butane by Temperature Modulation in Iron Doped Tin Oxide Sensors. *Sens. Actuators, B* **2006***, 115*, 610–613. DOI: 10.1016/j.snb.2005.10.046

[28] Zhang, G.; Xie, C. A Novel Method in the Gas Identification by Using $WO_3$ Gas Sensor Based on the Temperature-Programmed Technique. *Sens. Actuators, B* **2015**, *206*, 220–229. DOI: 10.1016/j.snb.2014.09.063

[29] Zhang, S.; Xie, C.; Hu, M.; Li, H.; Bai, Z.; Zeng, D. An Entire Feature Extraction Method of Metal Oxide Gas Sensors. *Sens. Actuators B,* **2008**, *132*(1), 81–89. DOI: 10.1016/j.snb.2008.01.015

[30] Burgués, J.; Marco, S. Feature Extraction for Transient Chemical Sensor Signals in Response to Turbulent Plumes: Application to Chemical Source Distance Prediction. *Sens. Actuators, B* **2020**, *320*, 128235. DOI: 10.1016/j.snb.2020.128235

[31] Acharyya, S.; Nag, S.; Guha, P. K. Ultra-Selective Tin Oxide-Based Chemiresistive Gas Sensor Employing Signal Transform and Machine Learning Techniques. *Anal. Chim. Acta* **2022**, *1217*, 339996. DOI: 10.1016/j.aca.2022.339996

[32] Potyrailo, R. A.; Go, S.; Sexton, D.; Li, X.; Alkadi, N.; Kolmakov, A.; Amm, B.; St-Pierre, R.; Scherer, B.; Nayeri, M.; Wu, G.; Collazo-Davila, C.; Forman, D.; Calvert, C.; Mack, C.; McConnell, P. Extraordinary Performance of Semiconducting Metal Oxide Gas Sensors Using Dielectric Excitation. *Nature Electronics* **2020**, *3*(5), 280–289. DOI: 10.1038/s41928-020-0402-3

[33] Zaromb, S.; Stetter, J. R. Theoretical Basis for Identification and Measurement of Air Contaminants Using an Array of Sensors Having Partly Overlapping Selectivities. *Sens. Actuators* **1984**, *6*(4), 225–243. DOI: 10.1016/0250-6874(84)85019-2

[34] Kim, T.; Cho, W.; Kim, B.; Yeom, J.; Kwon, Y. M.; Baik, J. M.; Kim, J. J.; Shin, H. Batch Nanofabrication of Suspended Single 1D Nanoheaters for Ultralow-Power Metal Oxide Semiconductor-Based Gas Sensors. *Small* **2022**, *18*, 2204078. DOI: 10.1002/smll.20220407

[35] Cho, W.; Kim, T.; Shin, H. Thermal Conductivity Detector (TCD)-Type Gas Sensor Based on a Batch-Fabricated 1D Nanoheater for Ultra-Low Power Consumption. *Sens. Actuators, B* **2022**, *371*, 132541. DOI: 10.1016/j.snb.2022.132541

[36] Lim, Y.; Heo, J. I.; Madou, M.; Shin, H. Monolithic Carbon Structures Including Suspended Single Nanowires and Nanomeshes as a Sensor Platform. *Nanoscale Res. Lett.* **2013**, *8*, 1–9. DOI: 10.1186/1556-276X-8-492

[37] Hong, J.; Kim, B.; Shin, H. Mixed-Scale Poly (methyl methacrylate) Channel Network-Based Single-Particle Manipulation via Diffusiophoresis. *Nanoscale* **2018**, *10*(30), 14421–14431. DOI: 10.1039/c7nr07669j

[38] Lim, Y.; Chu, J. H.; Kwon, S. Y.; Shin, H. Increase in Graphitization and Electrical Conductivity of Glassy Carbon Nanowires by Rapid Thermal Annealing. *J. Alloys Compd.* **2017**, *702*, 465–471. DOI: 10.1016/j.jallcom.2017.01.098

[39] Paska, Y.; Stelzner, T.; Assad, O.; Tisch, U.; Christiansen, S.; Haick, H. Molecular Gating of Silicon Nanowire Field-Effect Transistors with Nonpolar Analytes. *ACS Nano* **2012**, *6*(1), 335−345.

[40] Dai, T.; Meng, G.; Deng, Z.; Chen, Y.; Liu, H.; Li, L.; Wang, S.; Chang, J.; Xu, P.; Li, X.; Fang, X. Generic Approach to Boost the Sensitivity of Metal Oxide Sensors by Decoupling the Surface Charge Exchange and Resistance Reading Process. *ACS Appl. Mater. Interfaces* **2020**, *12*(33), 37295–37304. DOI: 10.1021/acsami.0c07626

[41] Lei, T.; Rao, Z.; Zhang, S.; Cai, S.; Xie, C. The Irreversible R-T Curves of Metal Oxide Gas Sensor under Programmed Temperature Cycle. *Sens. Actuators, B* **2016**, *235*, 481−491.

[42] Yamazoe, N.; Shimanoe, K. Theory of Power Laws for Semiconductor Gas Sensors. *Sens. Actuators, B* **2008**, *128*(2), 566-573.

[43] Zhang, S.; Xie, C.; Zhang, G.; Zhu, Q.; Zhang, S. Synchronously Deriving Electron Concentration and Mobility by Temperature-And Oxygen-Dependent Conductivity of Porous ZnO Nanocrystalline Film. *J. Phys. Chem. C* **2015**, *119*(1), 695–702.

[44] Hua, Z.; Li, Y.; Zeng, Y.; Wu, Y. A Theoretical Investigation of the Power-Law Response of Metal Oxide Semiconductor Gas Sensors I: Schottky Barrier Control. *Sens. Actuators, B* **2018**, *255*, 1911–1919.

[45] Suematsu, K.; Oyama, T.; Mizukami, W.; Hiroyama, Y.; Watanabe, K.; Shimanoe, K. Selective Detection of Toluene Using Pulse-Driven $SnO_2$ Micro Gas Sensors. *ACS Appl. Electron. Mater.* **2020**, *2*(9), 2913–2920. DOI: 10.1021/acsaelm.0c00547

[46] Jung, G.; Hong, Y.; Hong, S.; Jang, D.; Jeong, Y.; Shin, W.; Park, J.; Kim, D.; Jeong, C. B.; Kim, D. U.; Chang, K. S.; Lee, J. H. A Low-Power Embedded Poly-Si Micro-Heater for Gas Sensor Platform Based on a FET Transducer and Its Application for $NO_2$ Sensing. *Sens. Actuators, B* **2021**, *334*, 129642. DOI: 10.1016/j.snb.2021.129642

[47] Chowdhury, S.; Heinis, T.; Grimsrud, E. P.; Kebarle, P. Entropy Changes and Electron Affinities from Gas-Phase Electron-Transfer Equilibria: A- + B = A + B-. *J. Phys. Chem.* **1986**, *90*(12), 2747–2752. DOI: 10.1021/j100403a037

[48] Hughes, B. M.; Lifshitz, C.; Tiernan, T. O. Electron Affinities from Endothermic Negative-Ion Charge-Transfer Reactions. III. NO, $NO_2$, $SO_2$, $CS_2$, $Cl_2$, $Br_2$, $I_2$, and $C_2H$. *J. Chem. Phys.* **1973**, *59*(6), 3162–3181. DOI: 10.1063/1.1680458

[49] Travers, M. J.; Cowles, D. C.; Ellison, G. B. Reinvestigation of the Electron Affinities of $O_2$ and NO. *Chem. Phys. Lett.* **1989**, *164*(5), 449–455. DOI: 10.1016/0009-2614(89)85237-6

[50] Liu, H.; Zhou, J.; Yu, L.; Wang, Q.; Liu, B.; Li, P.; Zhang, Y. High-Sensitivity $SO_2$ Gas Sensor Based on Noble Metal Doped $WO_3$ Nanomaterials. *Int. J. Electrochem. Sci.* **2021**, *16*(12), 211240. DOI: 10.20964/2021.12.39

[51] Nandy, T.; Coutu Jr, R. A.; Ababei, C. Carbon Monoxide Sensing Technologies for Next-Generation Cyber-Physical Systems. *Sensors* **2018**, *18*(10), 3443. DOI: 10.3390/s18103443

[52] Mineo, G.; Moulaee, K.; Neri, G.; Mirabella, S.; Bruno, E. $H_2$ Detection Mechanism in Chemoresistive Sensor Based on Low-Cost Synthesized $WO_3$ Nanorods. *Sens. Actuators, B* **2021**, *348*, 130704. DOI: 0.1016/j.snb.2021.130704

[53] Celotta, R. J.; Bennett, R. A.; Hall, J. L. Laser Photodetachment Determination of the Electron Affinities of OH, $NH_2$, NH, $SO_2$, and $S_2$. *J. Chem. Phys.* **1974**, *60*(5), 1740–1745. DOI: 10.1063/1.1681268

[54] Rothe, E. W.; Tang, S. Y.; Reck, G. P. Measurement of Electron Affinities of $O_3$, $SO_2$, and $SO_3$ by Collisional Ionization. *J. Chem. Phys.* **1975**, *62*(9), 3829–3831. DOI: 10.1063/1.430941

[55] Refaey, K. M.; Franklin, J. L. Endoergic Ion–Molecule-Collision Processes of Negative Ions. I. Collision of $I^-$ on $SO_2$. *J. Chem. Phys.* **1976**, *65*(5), 1994–2001. DOI: 10.1063/1.433298

[56] Refaey, K. M.; Franklin, J. L. Endoergic Ion-Molecule-Collision Processes of Negative Ions. III. Collisions of I$^-$ on $O_2$, CO and $CO_2$, *Int. J. Mass Spectrom. Ion Phys.* **1976**, *20*(1), 19–32. DOI: 10.1016/0020-7381(76)80029-0

[57] McWeeny, R. The Electron Affinity of $H_2$: a Valence Bond Study. *J. Mol. Struct.: THEOCHEM* **1992**, *261*, 403–413. DOI: 10.1016/0166-1280(92)87089-I

[58] Li, Z.; Yu, J.; Dong, D.; Yao, G.; Wei, G.; He, A.; Wu, H.; Zhu, H.; Huang, Z.; Tang, Z. E-Nose Based on a High-Integrated and Low-Power Metal Oxide Gas Sensor Array. *Sens. Actuators, B* **2023**, *380*, 133289. DOI: 10.1016/j.snb.2023.133289

[59] Thai, N. X.; Tonezzer, M.; Masera, L.; Nguyen, H.; Van Duy, N.; Hoa, N. D. Multi Gas Sensors Using One Nanomaterial, Temperature Gradient, and Machine Learning Algorithms for Discrimination of Gases and Their Concentration. *Anal. Chim. Acta* **2020**, *1124*, 85−93. DOI: 10.1016/j.aca.2020.05.015

[60] Kwon, D.; Jung, G.; Shin, W.; Jeong, Y.; Hong, S.; Oh, S.; Kim, J.; Bae, J. H.; Park, B. G.; Lee, J. H. Efficient Fusion of Spiking Neural Networks and FET-Type Gas Sensors for a Fast and Reliable Artificial Olfactory System. *Sens. Actuators, B* **2021**, *345*, 130419. DOI: 10.1016/j.snb.2021.130419

[61] Kang, M.; Cho, I.; Park, J.; Jeong, J.; Lee, K.; Lee, B.; Del Orbe Henriquez, D.; Yoon, K.; Park, I. High Accuracy Real-Time Multi-Gas Identification by a Batch-Uniform Gas Sensor Array and Deep Learning Algorithm. *ACS Sens.* **2022**, *7*, 430−440. DOI: 10.1021/acssensors.1c01204

[62] Lee, K.; Cho, I.; Kang, M.; Jeong, J.; Choi, M.; Woo, K. Y.; Yoon, K. J.; Cho, Y. H.; Park, I. Ultra-Low-Power E-Nose System Based on Multi-Micro-LED-Integrated, Nanostructured Gas Sensors and Deep Learning. *ACS Nano* **2022**, *17*(1), 539–551. DOI: 10.1021/acsnano.2c09314

[63] Cho, I.; Lee, K.; Sim, Y. C.; Jeong, J. S.; Cho, M.; Jung, H.; Kang, M.; Cho, Y. H.; Ha, S. C.; Yoon, K. J.; Park, I. Deep-Learning-Based Gas Identification by Time-Variant Illumination of a Single Micro-LED-Embedded Gas Sensor. *Light: Sci. Appl.* **2023**, *12*(1), 95. DOI: 10.1038/s41377-023-01120-7

[64] Rusyn, B.; Lutsyk, O.; Kosarevych, R.; Maksymyuk, T.; Gazda, J. Features Extraction from Multi-Spectral Remote Sensing Images Based on Multi-Threshold Binarization. *Sci. Rep.* **2023**, *13*(1), 19655. DOI: 10.1038/s41598-023-46785-7

[65] Steyerberg, E. W.; Harrell Jr, F. E.; Borsboom, G. J.; Eijkemans, M. J. C.; Vergouwe, Y.; Habbema, J. D. F. Internal Validation of Predictive Models: Efficiency of Some Procedures for Logistic Regression Analysis. *J. Clin. Epidemiol.* **2001**, *54*(8), 774–781. DOI: 10.1016/S0895-4356(01)00341-9

**Supporting Information**

# Ultralow-Power Single-Sensor-Based E-Nose System Powered by Duty Cycling and Deep Learning for Real-Time Gas Identification

*Taejung Kim[‡, 1], Yonggi Kim[‡, 2], Wootaek Cho[1], Jong-Hyun Kwak[1], Jeonghoon Cho[2], Youjang Pyeon[2], Jae Joon Kim[2,*], Heungjoo Shin[1,*].*

[1]Department of Mechanical Engineering, Ulsan National Institute of Science and Technology (UNIST), Ulsan, 44919, Republic of Korea

[2]Department of Electrical Engineering, Ulsan National Institute of Science and Technology (UNIST), Ulsan, 44919, Republic of Korea

[‡] These authors contributed equally

*Corresponding Author Email: J. J. Kim (jaejoon@unist.ac.kr), H. Shin (hjshin@unist.ac.kr)

# Supplementary Note

**Convolutional Neural Network (CNN) for Image Processing**

Convolutional neural networks (CNNs) are algorithms used for processing high-dimensional data, such as in image recognition. CNNs exhibit exceptional performance in recognizing local patterns and features within images. Through the convolution and pooling layers, small portions of images are investigated, features are detected, and overall image characteristics are learned.[1] This enables CNNs to effectively recognize features, even when the position of objects changes within the image. The convolution and pooling processes maintain the spatial information and structural layout of images while extracting features.[2] Consequently, CNNs excel in tasks such as image classification, object detection, and segmentation. They are also effective in preventing overfitting. Techniques such as dropout are employed in CNNs to prevent overfitting, where random neurons are deactivated during training, enhancing generalization performance.[2] The feature extraction process mainly involves two key techniques: First, the convolution process applies filters to images to detect features. Second, the pooling process, conducted after convolution, extracts features and reduces the map size to attain lower computational complexity.[3]

**References**

1   Yamashita, R.; Nishio, M.; Do, R. K. G.; Togashi, K. Convolutional neural networks: an overview and application in radiology. Insights into imaging 2018, 9, 611–629. DOI: 10.1007/s13244-018-0639-9

2   Alzubaidi, L.; Zhang, J.; Humaidi, A. J.; Al-Dujaili, A.; Duan, Y.; Al-Shamma, O.; Santamaría, J.; Fadhel, M. A.; Al-Amidie, M.; Farhan, L. Review of deep learning: Concepts, CNN architectures, challenges, applications, future directions. Journal of big Data 2021, 8, 1–74. DOI: 10.1186/s40537-021-00444-8

3   Liu, P.; Zhang, H.; Lian, W.; Zuo, W. Multi-level wavelet convolutional neural networks. *IEEE Access* **2019**, *7*, 74973–74985. DOI: 10.1109/ACCESS.2019.2921451

# Supplementary figures

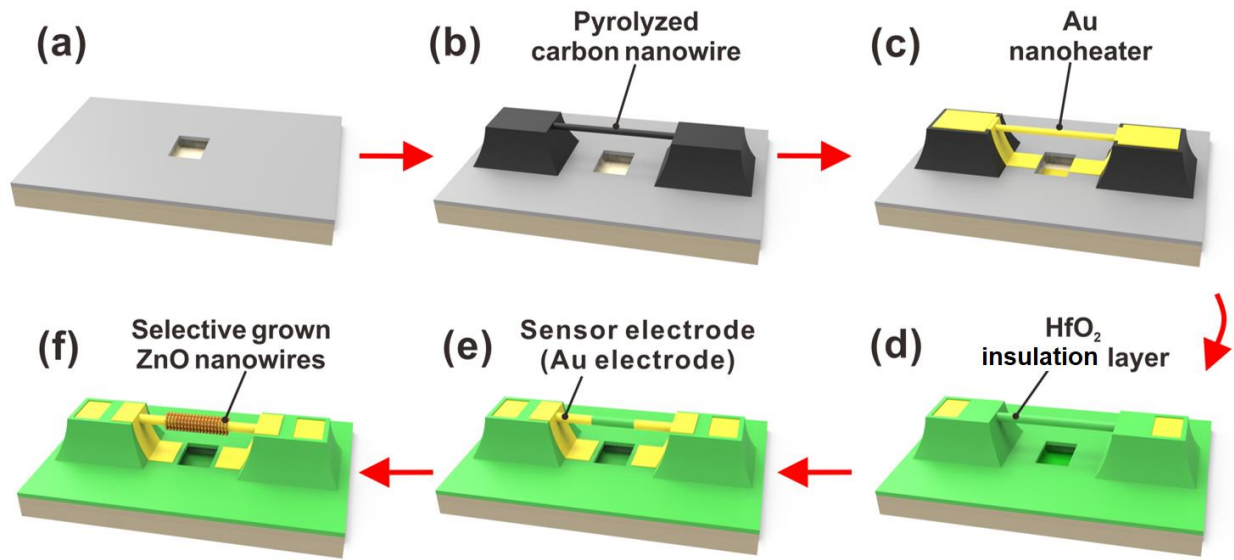

**Figure S1**. Schematic of the fabrication steps of a metal-oxide nanowire-based gas sensor with an embedded nanoheater: (a) Formation of the built-in shadow mask through isotropic silicon etching under a $SiO_2$ etch mask; (b) fabrication of a suspended carbon nanowire backbone; (c) integration of the nanoheater *via* selective Au layer deposition; (d) deposition and patterning of a $HfO_2$ insulation layer; (e) patterning of sensor electrodes (Au) for measuring the electrical resistance of the localized ZnO NW network; (f) localized growth of ZnO NWs.

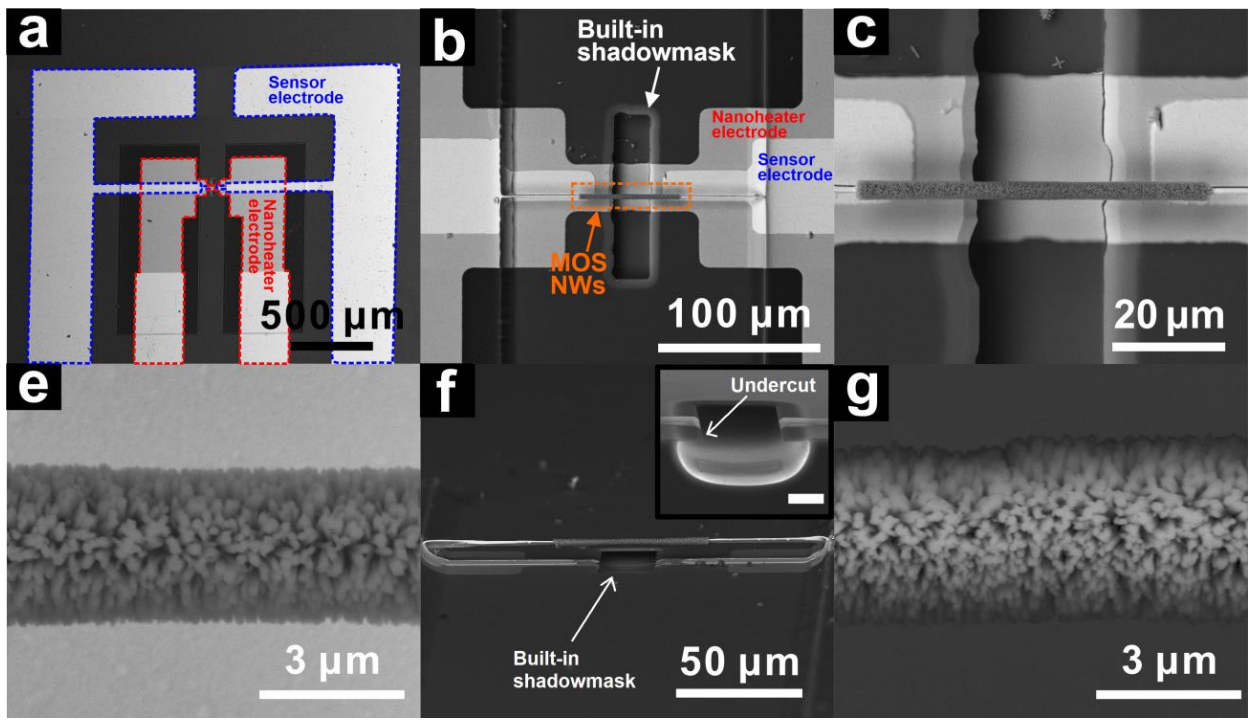

**Figure S2**. Scanning electron microscopy images of a suspended metal-oxide nanowire-based gas sensor with an embedded nanoheater: (a) Overall sensor configuration; (b–e) corresponding enlarged top views; (f, g) bird's-eye views of the suspended 1D structure (inset in (f): enlarged view of the built-in shadow mask, scale bar: 10 μm).

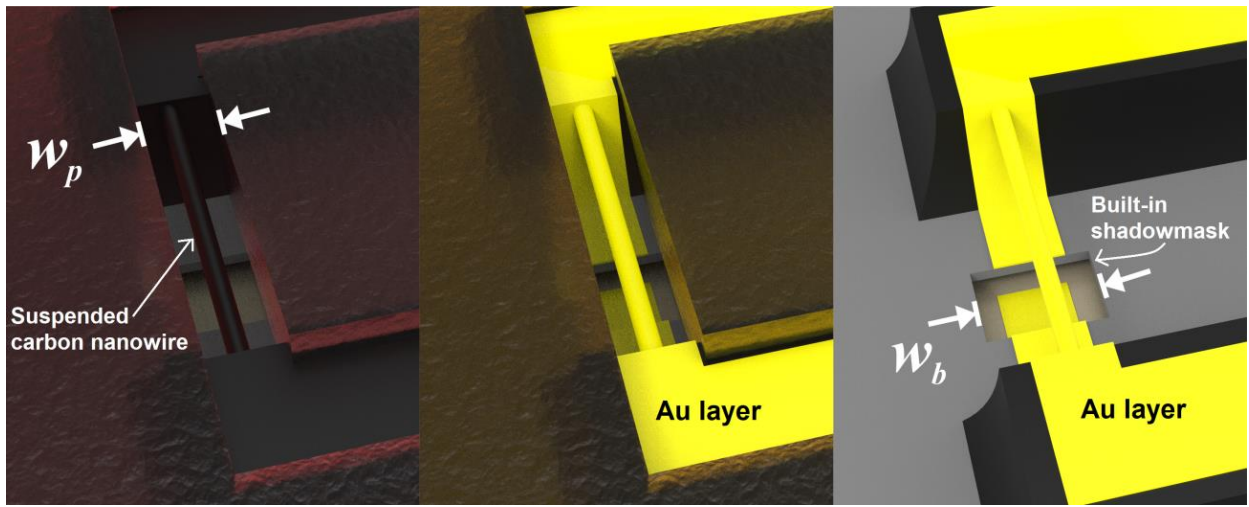

**Figure S3**. Schematic of selective metal coating on a 1D carbon nanostructure using a built-in shadow mask and an evaporation method: patterned photoresist mask over the carbon nanowire and built-in shadow mask (a) before and (b) after metal deposition; (c) the selectively metal-coated carbon nanowire following photoresist removal ($w_p$: width of the photoresist mask, $w_b$: width of the built-in shadow mask).

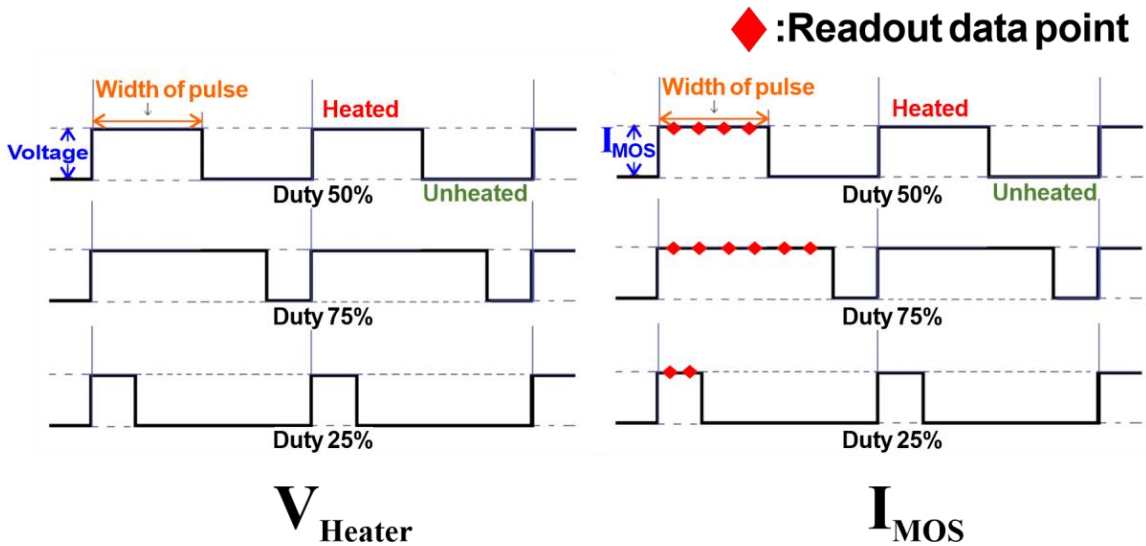

**Figure S4**. Diagrams illustrating the dependency of the sensor signals on the number of readout points in duty-cycling mode. The left diagrams show the heater's pulsed-voltage input ($V_{Heater}$), and the right diagrams display the corresponding sensor signal ($I_{MOS}$). Red diamonds represent readout data points.

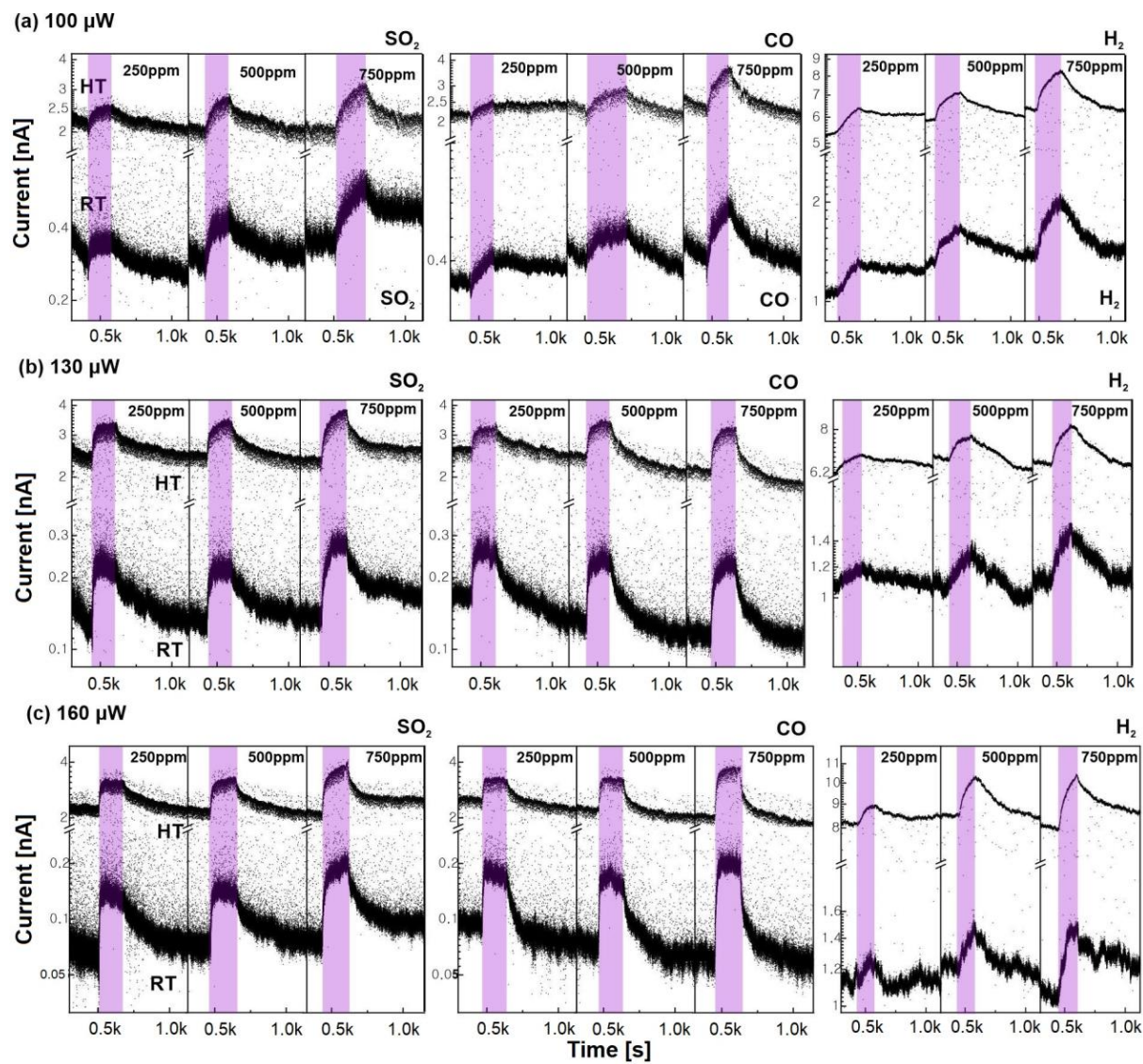

**Figure S5**. Transient dual-response current signals measured at various concentrations of reducing gases, including $SO_2$, CO, and $H_2$, under different power conditions ranging from 100 to 160 μW.

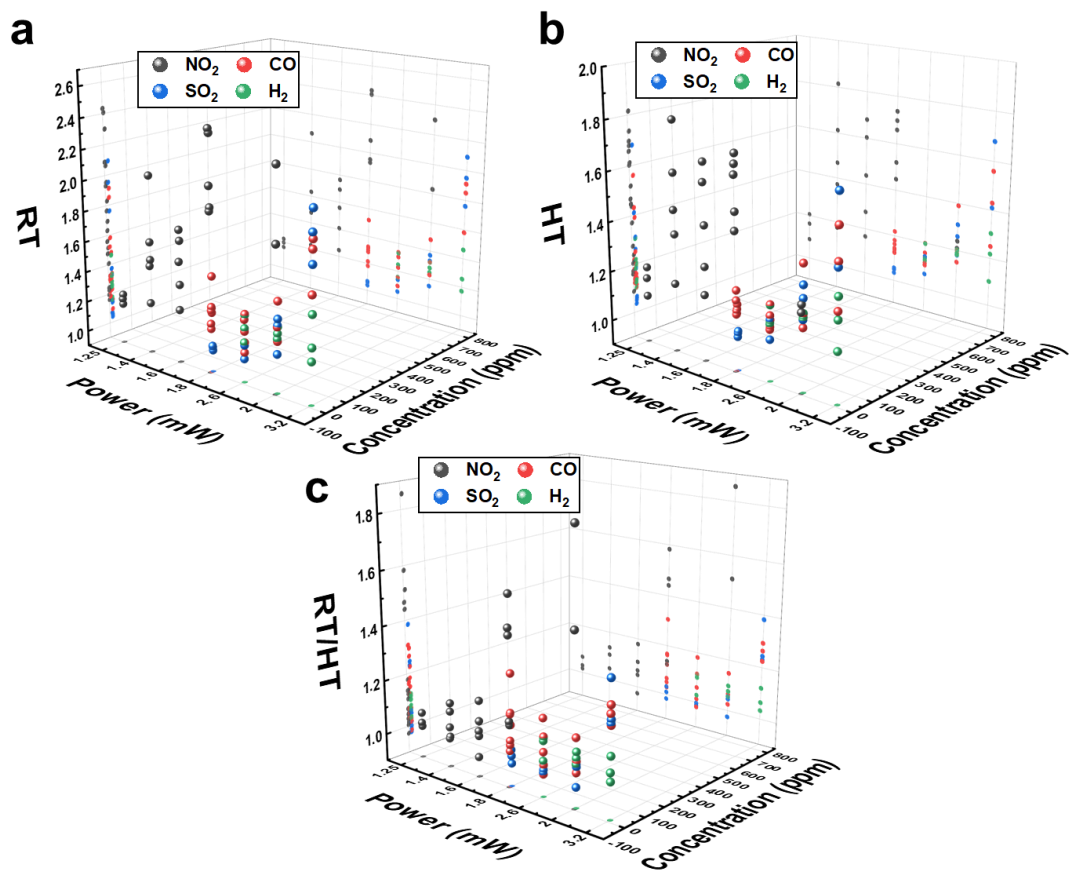

**Figure S6**. Comparison of gas (NO$_2$, CO, SO$_2$, and H$_2$) responses at (a) HT, (b) RT, and (c) their ratio (RT/HT) with respect to power consumption and concentration.

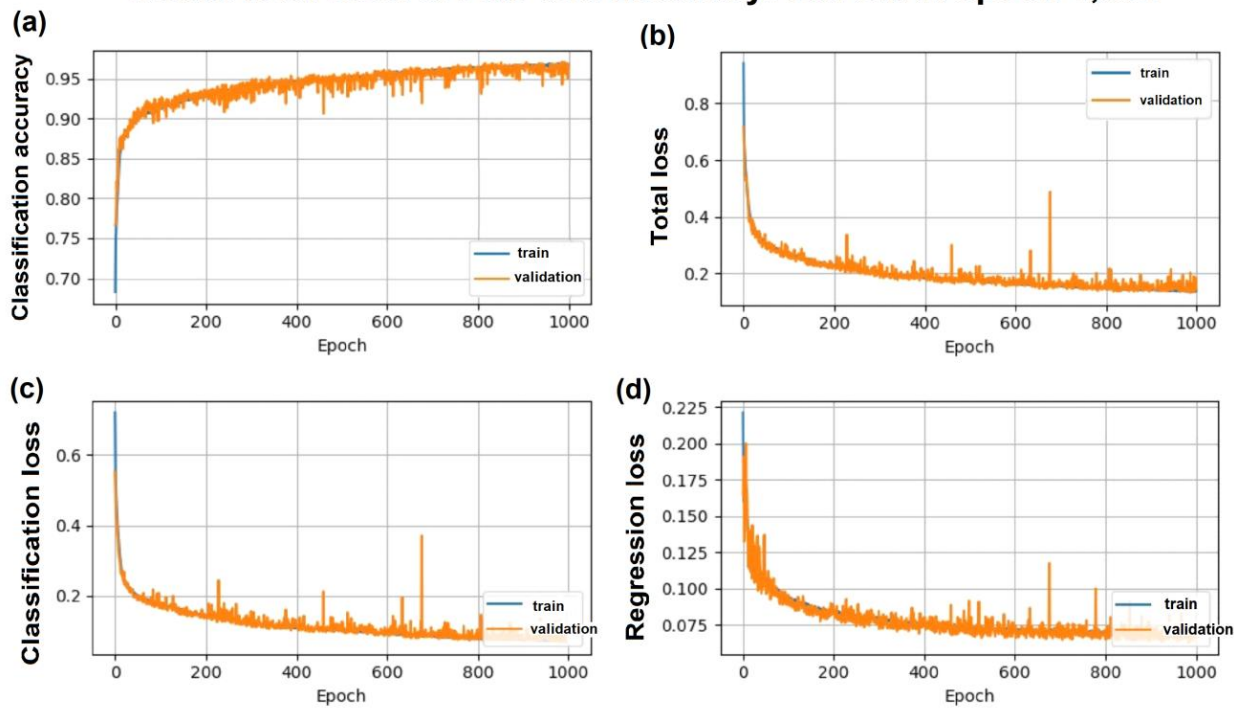

**Figure S7**. Evaluation of the effectiveness of gas identification using dual-response signals from a single MOS sensor through a CNN algorithm: (a) classification accuracy, (b) total loss, (c) classification loss, and (d) regression loss according to the number of epochs.

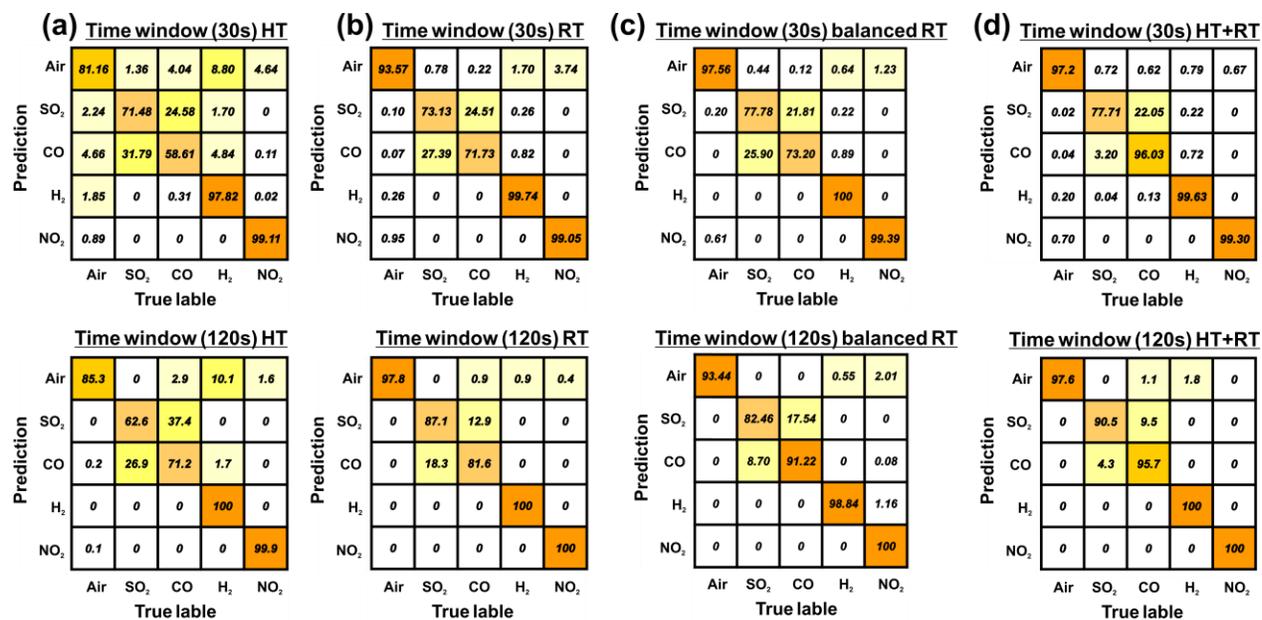

**Figure S8**. Confusion matrices of gas type identification for two time windows (30 and 120 s) obtained from (a) HT, (b) RT, (c) balanced RT, and (d) HT+RT data sets.

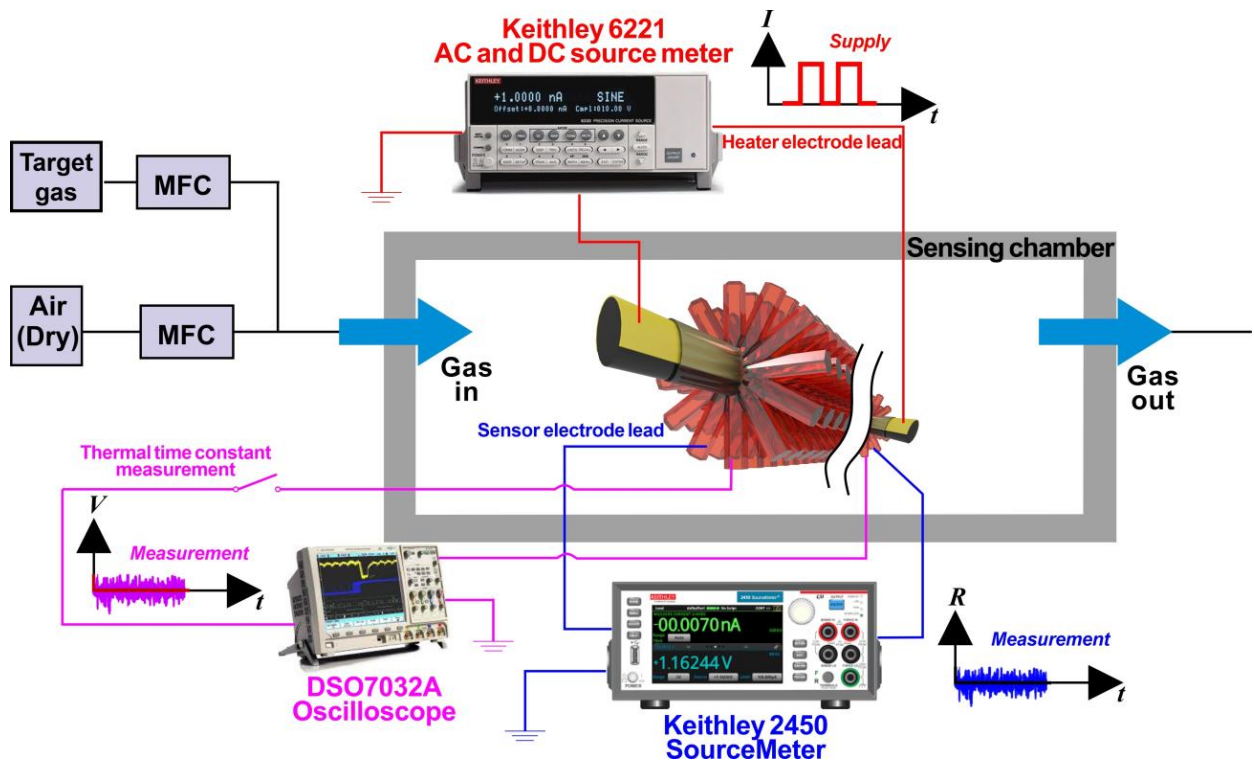

**Figure S9**. Gas sensing test setup.

**Table S1.** Comparison of the proposed and state-of-the-art duty-cycled MOS gas sensors.

| Sensor type | Heater size (W × L × T) [μm] | Thermal time constants (heating/cooling) [μs] | Duty cycle [%] | Power consumption [mW] | Reference |
|---|---|---|---|---|---|
| Chemiresistor | 20 × 30 × 1 | < 1000 (heating) | - | 2.51 | [14] |
| Chemiresistor | 20 × 50 × 0.46 | 76/33 | 16.6 | 2.1 **12.1 | [16] |
| Chemiresistor | 40 × 40 × 40 | - | 20 | - | [17] |
| Chemiresistor | 40 × 40 × 40 | - | 20 | - | [45] |
| FET | 2 × 38 × 1.1 | 115/60 | 50 | 1 **2 | [46] |
| Chemiresistor | 0.25 × 60 × 0.5 | 5/10 | 5 | 0.09 **1.8 | **This Work** |

**This value is calculated for the case of continuous heating.

**References** (same reference number as in main article)

[14] Choi, K. W.; Jo, M. S.; Lee, J. S.; Yoo, J. Y.; Yoon, J. B. Perfectly aligned, air-suspended nanowire array heater and its application in an always-on gas sensor. *Adv. Funct. Mater.* **2020**, *30*(39), 2004448. DOI: 10.1002/adfm.202004448

[16] Zhou, Q.; Sussman, A.; Chang, J.; Dong, J.; Zettl, A.; Mickelson, W. Fast response integrated MEMS microheaters for ultra low power gas detection. *Sens. Actuators, A* **2015**, *223*, 67–75. DOI: 10.1016/j.sna.2014.12.005

[17]   Suematsu, K.; Harano, W.; Oyama, T.; Shin, Y.; Watanabe, K.; Shimanoe, K. Pulse-driven semiconductor gas sensors toward ppt level toluene detection. *Anal. Chem.* **2018**, *90*(19), 11219–11223. DOI: 10.1021/acs.analchem.8b03076

[45]   Suematsu, K.; Oyama, T.; Mizukami, W.; Hiroyama, Y.; Watanabe, K.; Shimanoe, K. Selective detection of toluene using pulse-driven $SnO_2$ micro gas sensors. *ACS Appl. Electron. Mater.* **2020**, *2*(9), 2913–2920. DOI: 10.1021/acsaelm.0c00547

[46]   Jung, G.; Hong, Y.; Hong, S.; Jang, D.; Jeong, Y.; Shin, W.; Park, J.; Kim, D.; Jeong, C. B.; Kim, D. U.; Chang, K. S.; Lee, J. H. A low-power embedded poly-Si micro-heater for gas sensor platform based on a FET transducer and its application for $NO_2$ sensing. *Sens. Actuators, B* **2021**, *334*, 129642. DOI: 10.1016/j.snb.2021.129642